\documentclass[preprint2]{aastex63}

\usepackage{amsmath}
\usepackage{CJK}

\shorttitle{CMR in CNM}
\shortauthors{Kong et al.}

\begin{document}
\begin{CJK*}{UTF8}{gbsn}

\title{Filamentary Molecular Cloud Formation via Collision-induced Magnetic Reconnection in Cold Neutral Medium}
% \title{Magnetically Regulated Molecular Cloud and Star Formation -- CMR in CNM}

\author[0000-0002-8469-2029]{Shuo Kong (孔朔)}
\affiliation{Steward Observatory, University of Arizona, Tucson, AZ 85719, USA}

\author[0000-0002-0820-1814]{Rowan J. Smith}
\affiliation{SUPA, School of Physics and Astronomy, University of St Andrews, North Haugh, St Andrews, KY16 9SS}

\author[0000-0001-9956-0785]{David Whitworth}
\affiliation{Universidad Nacional Aut\'onoma de M\'exico, Instituto de Radioastronom\'ia y Astrof\'isica, Antigua Carretera a P\'atzcuaro 8701, Ex-Hda. \\
San Jos\'e de la Huerta, 58089 Morelia, Michoac\'an, M\'exico}

\author[0000-0002-3131-7372]{Erika T. Hamden}
\affiliation{Steward Observatory, University of Arizona, Tucson, AZ 85719, USA}

\begin{abstract}
We have investigated the possibility of molecular cloud formation 
via the Collision-induced Magnetic Reconnection (CMR) mechanism of the cold neutral medium (CNM).
Two atomic gas clouds with conditions typical of the CNM
were set to collide at the interface of reverse magnetic fields.
The cloud-cloud collision triggered magnetic reconnection and produced a giant 20\,pc 
filamentary structure which was not seen in the control models without CMR.
The cloud, with rich fiber-like sub-structures,
developed a fully molecular spine at 5\,Myr.
Radiative transfer modeling
of dust emission at far infrared wavelengths showed
that the middle part of the filament contained dense cores over a span of 5\,pc.
Some of the cores were actively forming stars and 
typically exhibited both connecting fibers in
dust emission and high-velocity gas in CO line emission,
indicative of active accretion through streamers.
Supersonic turbulence was present in and around the CMR-filament due to
inflowing gas moving at supersonic velocities in the collision mid-plane.
The shocked gas was condensed and transported to the main filament 
piece by piece by reconnected fields,
making the filament and star formation a bottom-up process.
Instead of forming a gravitationally bounded cloud
which then fragments hierarchically (top-down) and forms stars,
the CMR process creates dense gas pieces and magnetically transports 
them to the central axis to constitute the filament.
Since no turbulence is manually driven,
our results suggest that CMR is capable of self-generating turbulence.
Finally, the resulting helical field
should show field-reversal on both sides of the filament from
most viewing angles.
\end{abstract}


\section{Introduction}\label{sec:intro}
\end{CJK*}

%Stars form in molecular clouds. 
The formation and evolution of filamentary molecular clouds
are crucial for our understanding of star formation and the evolution
of the interstellar medium in galaxies \citep{2023ASPC..534..153H}. 
To put it simply, gas in the interstellar 
medium (ISM) contracts to form clouds when it becomes self-gravitating,
%A few mechanisms possibly trigger the contraction, e.g.,
%large-scale colliding flows and gravitational instability 
which could be triggered by mechanisms such as 
colliding flows and gravitational instability 
\citep[see][and references therein]{2023ASPC..534....1C}.
During this process gas may begin as cold neutral medium (CNM)
and transition to molecular while getting denser.

Observations have told us
that a significant number of clouds show a prominent filamentary
structure, e.g., the Orion A cloud as a whole
\citep{2013ApJ...777L..33P,2015A&A...577L...6S}, the
L1495 filament in the Taurus cloud
\citep{2012ApJ...756...12L,2013A&A...550A..38P,2019A&A...623A..16S}, 
and the California cloud as a whole
\citep{2009ApJ...703...52L,2021ApJ...908...76L}. 
Several models for filament formation have been proposed.
For instance, the
collapse of an initially asymmetric sheet can produce a
filamentary structure \citep{2004ApJ...616..288B}.
Supersonic turbulence can create shocked layers which may evolve
into filaments \citep{2004ApJ...609L..83L},
but these filaments should be stochastically 
distributed %while the overall cloud structure is still spherically symmetric, 
unless a significant source of anisotropy is present. 
Note, here we are 
referring to local mechanisms that produce filaments that
may or may not be parallel to the Galactic plane. Large-scale
filamentary clouds that are triggered by spiral potentials
in the Galactic disk are out of the scope of this work
\citep[but see, for example,][]{2020MNRAS.492.1594S}.

Helical magnetic fields provide a cylindrical symmetry that
can hold a filamentary structure, making it possible to
have isolated large filamentary clouds. For instance,
\citet{1997ApJS..111..245H} showed a large-scale
line-of-sight reverse magnetic field around the famous 
Orion A giant molecular cloud. 
% However, it is not clear if this 
% indicates a helical field around the cloud, or the projection
% of some other field configuration 
% \citep[see][]{2011ApJ...741..112P,2019A&A...632A..68T},
% although a helical interpretation was used to model the cloud dynamics
% \citep[e.g., see][]{2016A&A...590A...2S}.
While \citet{1997ApJS..111..245H,2019A&A...632A..68T}
favored a bow-shape field, \citet{2011ApJ...741..112P}
instead inferred a helical field wrapping the OMC-1 filament.
Theoretically,
\citet{2000MNRAS.311...85F} showed in their model that
a helical field stabilizes a cylindrical cloud. However,
such a field configuration was not previously seen in numerical simulations \citep[see, however, a recent work by][]{2024arXiv240518474Z}. 

In an effort to explain a peculiar observational feature
in the Stick filament in the Orion A cloud,
\citet[][hereafter K21]{2021ApJ...906...80K} proposed a 
collision-induced magnetic reconnection (CMR) mechanism that was 
able to form a single filament with helical fields. Their model
incorporated a field reversal pattern into the initial conditions 
(motivated by the Heiles observation in Orion)
and successfully reproduced the observed morphology, 
density distribution, and kinematics of the Stick.
Field-reversals are also observed around other
filamentary clouds, including the Perseus cloud and the
California cloud \citep{2018A&A...614A.100T,2022A&A...660A..97T}.
In the Galactic disk, we know there are large-scale field-reversals
\citep{2018ApJS..234...11H}. Any gas motion that brings
together the reverse field can trigger CMR, even if the
field is not exactly anti-aligned 
\citep[see Figures 26, 27 in][]{2021ApJ...906...80K}.

The helical field,
which was a natural result of CMR, created and maintained
the filament with relatively strong magnetic surface pressure.
It showed that magnetic fields can actively produce dense gas,
unlike the stereotype in which magnetic fields only passively
hinder gas concentration. Later, 
\citet[][hereafter K22]{2022MNRAS.517.4679K} 
showed that a CMR-filament like the Stick would 
collapse along its major axis and form 
a star cluster. However, the star formation rate was
relatively low due to the surface field that hindered
gas accretion (which was only possible through streamers).

If one sub-filament in the Orion A cloud is explicable by CMR,
then the natural follow-up question is whether the entire
Orion A formed in a similar fashion.
In this paper, we follow up K22 with
a pilot investigation
of the formation of a filamentary molecular cloud via CMR
in the cold neutral medium (CNM) rather than molecular gas.
Specifically, we aim to study the formation
of a 10-pc-scale filamentary cloud from CNM.
The goal is to see if CMR has the potential to form
a long molecular filament like the Orion A. For example, is it able
to form a filament 10 times longer by simply scaling up
the initial sizes in K22? Does it produce fibers
as seen in observations \citep[e.g.,][]{2018A&A...610A..77H}?
Is it capable of generating turbulence? Most importantly,
can CMR convert CNM to molecular gas?
At this point, we do not intend to reproduce any specific cloud,
but using Orion A as a guide for our exploration.
If the answer to the above questions is yes, meaning
the potential is there, then we can attempt to scale the initial
conditions accordingly to reproduce key features for specific
clouds. For example, although the CMR model in this paper does not
match the high column density of Orion A, we can
increase the initial density to produce a filament with higher
densities. Although we are forming a 10-pc filament,
we can increase the initial scales to produce a
100-pc filament just like Orion A \citep{2018A&A...619A.106G}.

In reality, the initial condition might not be the collision
between two spherical clouds. It could be the expansion of
a bubble that hits a wall of gas with reverse fields. The
bubble and the wall could have multiple pairs of clumps
that collide and trigger CMR, thus forming a
number of sub-filaments that as a whole resemble a large filament.
Nonetheless, the exploration in this paper provides
a useful initial look into the capability of CMR and points
us to the right direction in terms of modeling observed
filaments.

Although the original model of CMR in K21 started with
fully molecular conditions, the very beginning of 
a giant molecular cloud is likely CNM, if not warm neutral
medium (WNM). Observationally, this conjecture relies on
the fact that we have not clearly seen large-scale colliding 
molecular gas on either side of the Orion A filament
\citep[e.g.,][]{2018ApJS..236...25K}. On the other hand,
there are narrow velocity components of HI gas around
Orion A that appear to be CNM \citep{2018A&A...609L...3S}. Such an origin appears feasible and theoretically, CNM is the precursor of molecular gas.
%Theoretically, CNM is the precursor of molecular gas clouds
%in which star formation happens. Taking these into
%consideration, we believe it is crucial to investigate
%CMR in CNM.
Therefore, in this paper, we model the formation of
a molecular filament via CMR in CNM. We will examine the
aforementioned properties, including the molecular fraction,
the sub-structures of the filament, and (supersonic) turbulence.

\begin{figure*}[htb!]
\centering
\epsscale{1.1}
\plottwo{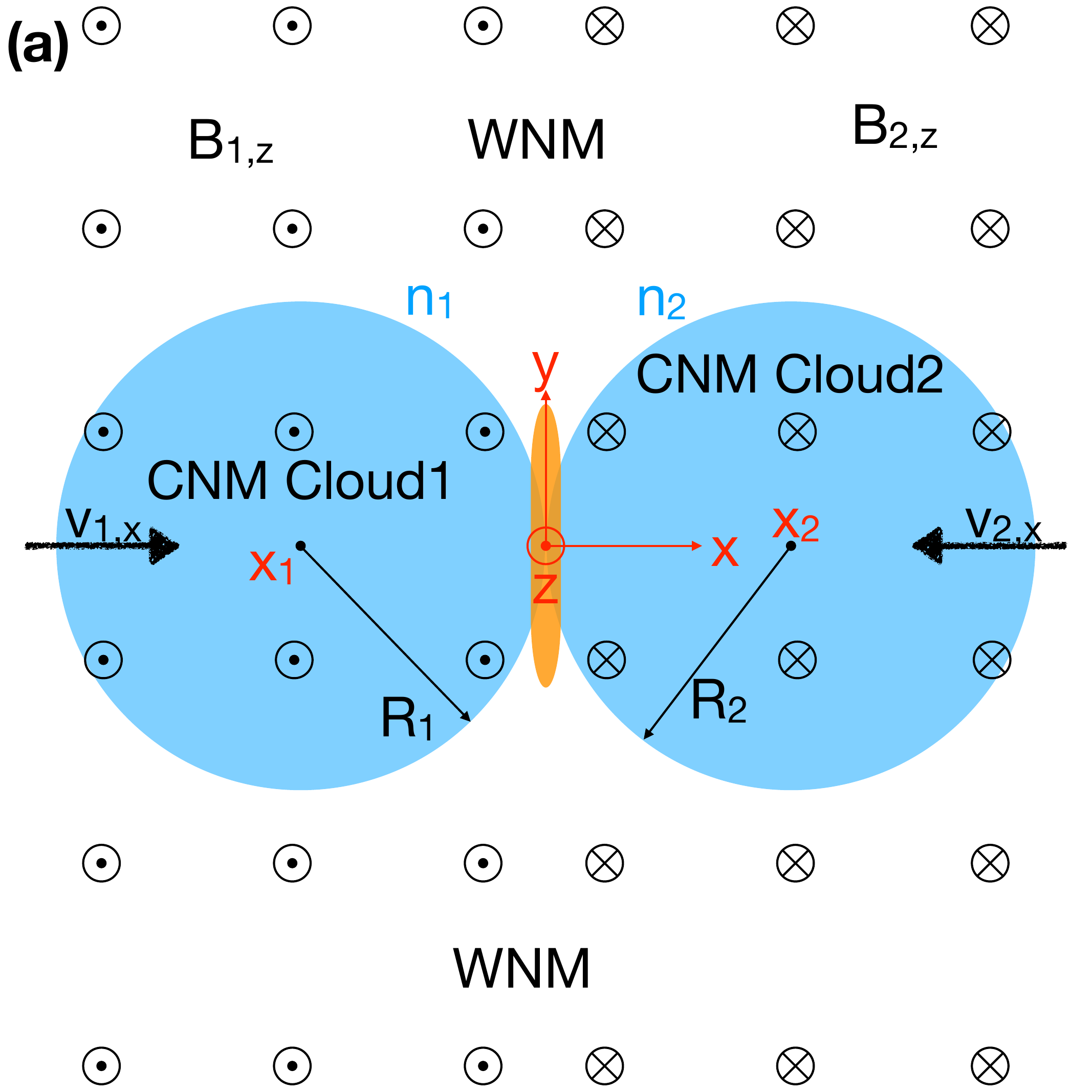}{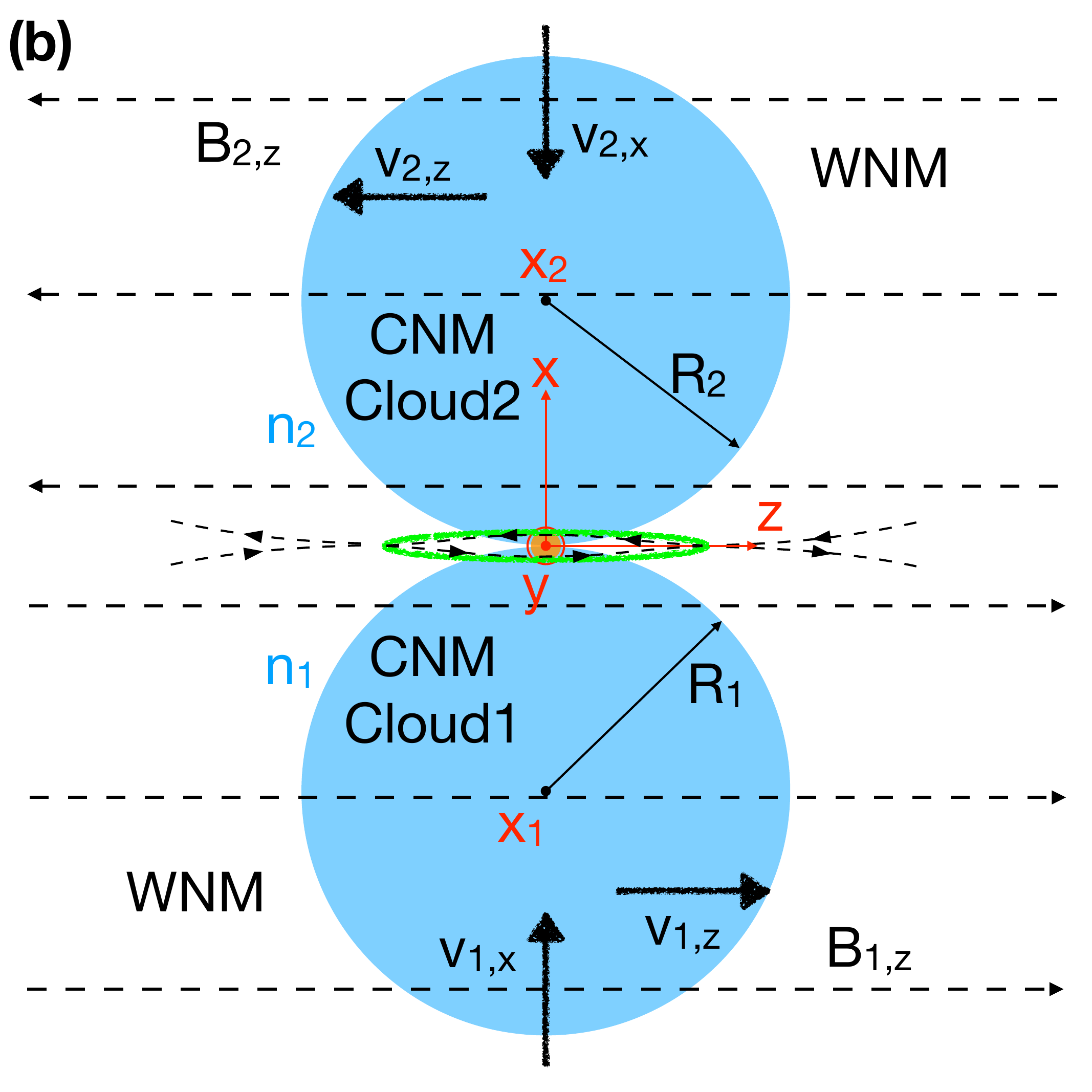}
\caption{
\mbox{MRCOLA\_CNM} initial condition in two viewing angles. {\bf (a):} A view in the x-y plane. Two spherical CNM clouds are in thermal pressure balance with the ambient WNM. They collide at the origin of the Cartesian coordinate system (red) in which the z-axis points toward us (red circle-point, right-hand rule). The colliding velocities are $v_{\rm 1,x}$ and $v_{\rm 2,x}$ for the two clouds, respectively. Magnetic fields point toward us (black circle-points) for $x<0$ and away from us (black circle-crosses) for $x>0$. The collision triggers the formation of the {\it filament} (orange) along the y-axis. Note the {\it filament} is imaginary in this initial condition figure. It will form after the cloud collision. {\bf (b):} A view in the z-x plane. In this view, magnetic fields are parallel to the plane of the sky. The y-axis points toward us (red circle-point) so we see the {\it filament} (orange) cross-section (again imaginary). The green ellipse marks the location of the compression {\it pancake} if no magnetic fields. With CMR, the field reconnects at two edges of the {\it pancake} and forms a {\it field-loop} (black dashed arrow-curve inside the green ellipse) around the {\it pancake}. The magnetic tension squeezes the {\it pancake} to the central axis (y-axis), making it a {\it filament}. Outside the {\it pancake}, the reconnected field forms two half loops and pulls gas away.
\label{fig:setup}}
\end{figure*}

In the following, we first briefly introduce our simulation
setup in \S\ref{sec:method}. Then, we report our findings 
in the results section \S\ref{sec:results}. We discuss
the implications based on the results in \S\ref{sec:discus}.
Finally, we summarize our findings and conclude in
\S\ref{sec:conclu}.

\section{Model Setup}\label{sec:method}

Following K22, we use a modified version of the \textsc{Arepo}
code \citep{Springel10} to simulate CMR. 
The code configuration is the same as K22, 
including magnetohydrodynamics \citep[MHD,][]{Pakmor11},
gravity \citep{Springel05},
time-dependent chemistry \citep{Gong17,Clark19} including gas
self-shielding \citep{Clark12b} from 
an ambient interstellar radiation field,
heating/cooling processes \citep{Clark19}, 
and sink particles to represent sites of star formation 
\citep{Bate95,Greif11,2020MNRAS.492.2973T}. 
A refinement that ensures the Jeans scale be resolved
by at least 16 cells is adopted \citep{Truelove97}.
See K22 method section for details.
% Since we are modeling a 
% factor of 10 larger scale, we adopt a different unit system
% from K22 with the length unit set to 10 pc,
% the time unit set to 20 Myr, 
% the mass density unit to $3.84\times10^{-23}$ g cm$^{-3}$
% (corresponding to $n_{\rm H,tot}$=17 cm$^{-3}$ assuming a 
% mean molecular mass per H of $\mu_H=1.4$).
% As a result, the velocity unit is still 0.51 km s$^{-1}$,
% the mass unit becomes 560 M$_\odot$,
% the magnetic field unit becomes 0.31 $\mu$G, and the 
% gravitational constant is $G=1$.

% However, in the ISM there are several additional possibility of how such anti-aligned fields could be generated. Supernovae explosions and feedback drive bubbles throughout the galactic ISM \citep[e.g.,][]{Padoan16,Watkins23}. Since magnetic fields lie preferentially parallel to isodensity contours \citep[e.g.,][]{Soler13, PlanckXIX} they will thread the walls of the bubbles. Several prominent local molecular clouds have been shown from 3D dust measurements to form at the boundaries of intersecting bubbles \citep{Zucker22}. Since the field on the right and left side of a bubble from any given angle will have opposite direction, it is likely that any collisions between gas in these shells will have opposite field directions similar to the case we consider here.

\begin{deluxetable*}{@{\extracolsep{4pt}}cccc}[htb!]
\tablecaption{Model Initial Conditions \label{tab:ic}}
\tablehead{
\colhead{Model} & \colhead{\mbox{MRCOLA\_CNM}} & \colhead{\mbox{MRCOLA\_CNM\_sameB}} & \colhead{\mbox{MRCOLA\_CNM\_noB}}\\
\colhead{(1)} & \colhead{(2) fiducial} & \colhead{(3)} & \colhead{(4)}
}
\startdata
$L$ & 80 pc & 80 pc & 80 pc \\
$T_{\rm CNM}$ & 100 K & 100 K & 100 K \\
$T_{\rm WNM}$ & 6000 K & 6000 K & 6000 K \\
$\zeta$ & $3.0\times10^{-17}$ s$^{-1}$ & $3.0\times10^{-17}$ s$^{-1}$ & $3.0\times10^{-17}$ s$^{-1}$ \\
ISRF & 1.7$G_0$ & 1.7$G_0$ & 1.7$G_0$ \\
$n_{\rm WNM}$ & 0.5 cm$^{-3}$ & 0.5 cm$^{-3}$ & 0.5 cm$^{-3}$ \\
\hline
$n_1$ & 30 cm$^{-3}$ & 30 cm$^{-3}$ & 30 cm$^{-3}$ \\
$R_1$ & 9 pc & 9 pc & 9 pc \\
$v_{\rm 1,x}$ & 2.5 km s$^{-1}$ & 2.5 km s$^{-1}$ & 2.5 km s$^{-1}$ \\
$v_{\rm 1,z}$ & 0 & 0 & 0 \\
$B_{\rm 1,z}$ & 5 $\mu$G & 5 $\mu$G & 0 \\
\hline
$n_2$ & 30 cm$^{-3}$ & 30 cm$^{-3}$ & 30 cm$^{-3}$ \\
$R_2$ & 9 pc & 9 pc & 9 pc \\
$v_{\rm 2,x}$ & -2.5 km s$^{-1}$ & -2.5 km s$^{-1}$ & -2.5 km s$^{-1}$ \\
$v_{\rm 2,z}$ & 0 & 0 & 0 \\
$B_{\rm 2,z}$ & -5 $\mu$G & 5 $\mu$G & 0
\enddata
\tablecomments{The physical quantities follow the notation in K22 Figure 1. $L$ is the domain size. $T_{\rm CNM}$ is the initial temperature of the CNM clouds. $T_{\rm WNM}$ is the ambient WNM temperature. $\zeta$ is the cosmic-ray ionization rate. ISRF is in unit of Habing field $G_0$. $n_{\rm WNM}$ is the ambient WNM number density. $n_1$ is the Cloud1 H number density. $R_1$ is the Cloud1 radius. $v_{\rm 1,x}$ is the Cloud1 collision velocity. $v_{\rm 1,z}$ is the Cloud1 shear velocity. $B_{\rm 1,z}$ is the B-field for x$<$0. $n_2$ is the Cloud2 H number density. $R_2$ is the Cloud2 radius. $v_{\rm 2,x}$ is the Cloud2 collision velocity. $v_{\rm 2,z}$ is the Cloud2 shear velocity. $B_{\rm 2,z}$ is the B-field for x$>$0.}
\end{deluxetable*}

Here we will use as an initial condition a simple toy model of two colliding spherical clouds. Figure \ref{fig:setup} shows the initial conditions for the fiducial model.
Hereafter we name the fiducial model \mbox{MRCOLA\_CNM} (as opposed 
to the fiducial model \mbox{MRCOLA} in K22).
The initial condition for \mbox{MRCOLA\_CNM}
is structurally the same as \mbox{MRCOLA} in that
two spherical clouds are embedded in a less dense medium and
collide at the interface of anti-parallel (reverse) magnetic fields.
Unlike in K22, here the colliding clouds are CNM that is embedded in WNM. 
Accordingly, the initial gas composition is fully atomic
instead of molecular. The elemental abundances
are the same as those in K22 and correspond to solar values.
Since we are modeling the scale of 10 pc,
% \footnote{The total length of Orion A can be much larger \citep{2018A&A...619A.106G}. Our objective is not reproducing Orion A but exploring the potential of molecular cloud formation via CMR at 10 pc scales, which is more relevant to the scale of the integral-shaped filament \citep{1987ApJ...312L..45B}. For Orion A, we suspect that it was not a result of CMR from the collision of only one pair of clouds. A more realistic scenario could be that multiple CMR-filaments together constituted the large filament, which is beyond the scope of this paper.},
we enlarge simulation scales by a factor of 10 (compared to K22).
Specifically, the colliding clouds now have radii of 9 pc 
and the domain has a size of 80 pc.
The relatively large domain size is to avoid boundary artifacts. 

\begin{figure*}[htb!]
\centering
\epsscale{1.1}
\plotone{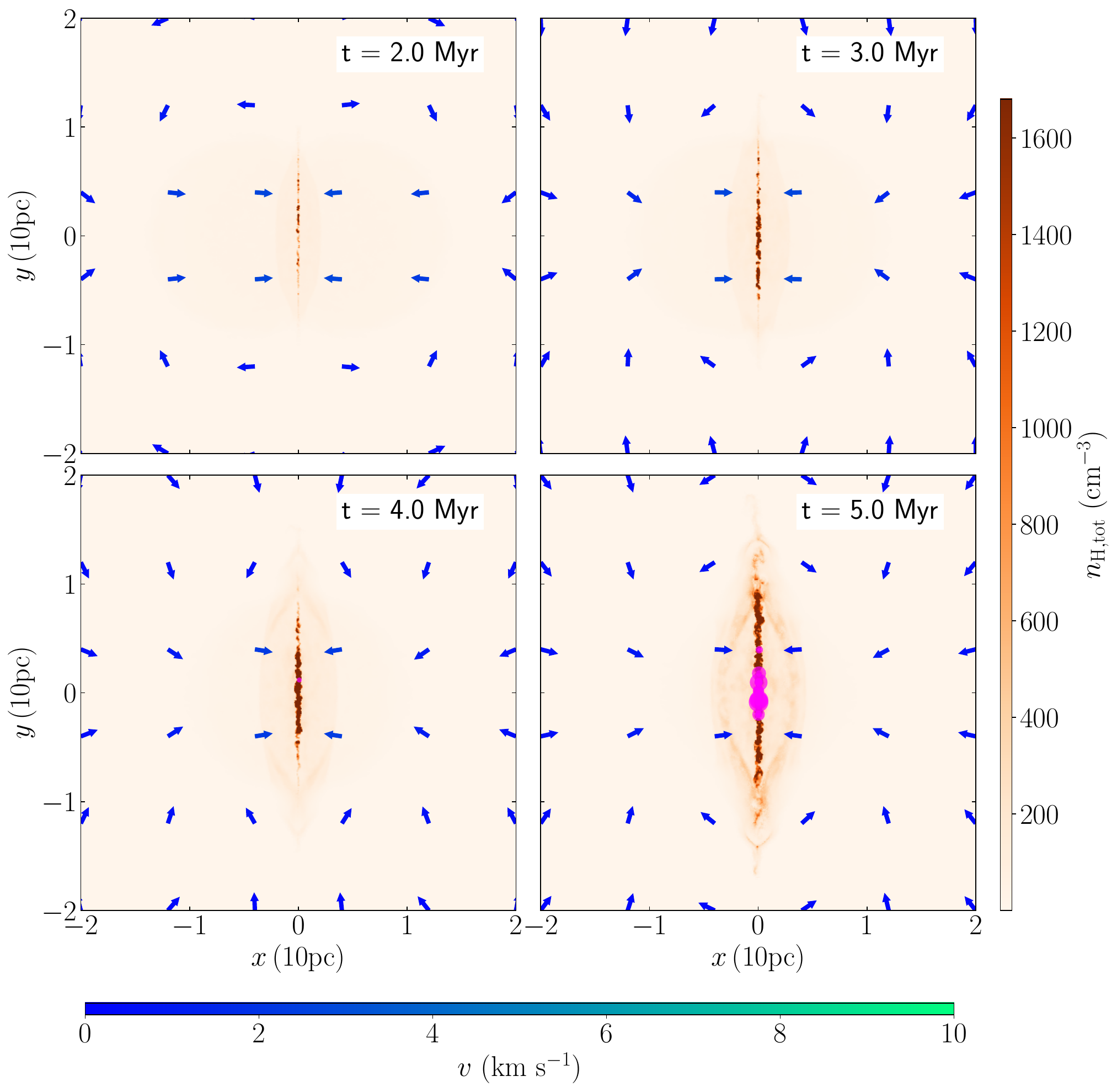}
\caption{
\mbox{MRCOLA\_CNM} density slice plots for the z=0 plane. 
The linear color scale shows the total gas density in unit of $n_{\rm H,tot}$.
The magenta filled circles show the sinks. The circle size is proportional to
the sink mass. The time step is shown at the top-right in each panel.
The arrows show the velocity vectors. The arrow color scales with the
velocity magnitude (the bottom color bar).
This slice corresponds to Figure \ref{fig:setup}(a).
\label{fig:zslice}}
\end{figure*}

For the clouds, we adopt typical physical properties for the
CNM, i.e., a total H number density 30 cm$^{-3}$ and a 
gas kinetic temperature $T_{\rm CNM}$ = 100 K.
For the ambient gas,
we adopt typical properties for WNM, i.e., $n_{\rm WNM}$ = 0.5 cm$^{-3}$ 
and $T_{\rm WNM}$ = 6000 K, so that it is in thermal pressure balance
with the clouds.
Each CNM cloud has a total mass of $\sim$3000 M$_\odot$.
The WNM total mass is $\sim$3 times the CNM total mass.
The relative collision speed ($\sim$5 km s$^{-1}$)
considers the typical relative speed between multiple line components
toward interstellar clouds. For instance, the HI observation by
\citet{1997ApJS..111..245H} showed multiple HI components around
Orion A with $\sim$5-10 km s$^{-1}$ separations. Around the Taurus
B213 filament, \citet{2019A&A...623A..16S} showed a similar separation
between HI components and a few km s$^{-1}$ separation in molecular gas
emission. We thus set the speed for each colliding CNM
cloud to 2.5 km s$^{-1}$. Unlike K22, we set the z-velocity to zero
so the collision is head-on. Adding the shear velocity or an impact
parameter will cause the rotation of the filament (see for example K22), 
which is not our focus in this paper. The magnetic field strength is set
to 5 $\mu$G in each direction, simply following the HI Zeeman observations 
in \citet{1997ApJS..111..245H}.
Table \ref{tab:ic} column (2) lists the fiducial model parameters. 
The physical quantities follow the notation in Figure \ref{fig:setup}.

\section{Results and Analysis}\label{sec:results}

In the following, we will be describing the physical
process in a mostly qualitative way to show the physical picture.
The goal is to test whether a molecular cloud can form from atomic
gas (CNM) through the CMR process. We resort to future papers for
detailed investigations of the various physical processes.

We define the nomenclature of a few most important
objects during the CMR process. From now on,
a {\it pancake} refers to the compressed circular
structure between the colliding spheres.
A {\it filament} refers to the main elongated
structure formed along the y-axis.
A {\it fiber} refers to
the small wiggly structures in and around the filament.
The {\it filament} can consist of numerous {\it fibers}.
A {\it ring} refers to the circular and/or elliptic
structure along the {\it filament}.
A {\it fork} refers to the open tuning-fork-like structure
at two ends of the {\it filament}. A {\it field-loop} refers to
the magnetic field loop around the {\it pancake} after magnetic
reconnection (the dashed loop around the {\it pancake} in
Figure \ref{fig:setup}(b)). All other names not in the above
nomenclature are considered temporary for description purpose.

CMR is necessarily a 3D process with complex dynamics due to
the interaction between gas and magnetic fields.
To help readers better understand the process,
we refer to a movie (file name {\tt CMRfield.mpg})
in \url{https://doi.org/10.7910/DVN/CXHWRR}
that shows the CMR process based on simulations from K21.
The movie shows the field lines and the gas rendering
near the field-reversal plane (the x=0 plane in Figure \ref{fig:setup}).
Our coordinate system is similar to their setup.
In the movie, one can see the {\it pancake} initially which
is later wrapped by {\it field-loops}, and the {\it filament}
which is wrapped by helical fields at the end.
The resulting field around the filament is illustrated
in Figure 2 of \citet{2023ApJS..265...58K}.

\subsection{Density Structures}\label{subsec:density}

Figure \ref{fig:zslice} shows density slice plots for z=0 plane.
This plane corresponds to the setup as shown in Figure 1(a).
Here, two spherical CNM clouds with a radius of 9 pc collide at x=0. 
Magnetic fields at x$<$0 point toward us and at x$>$0
point away from us. The x=0 plane is the field-reversal plane.
Note, the spherical geometry is not a necessary requirement for CMR.
Provided the colliding clouds 
have a protruding shape CMR-filament formation should be triggered
\citep[K21,][]{2022ApJ...933...40K}.

Initially, the two CNM clouds move toward the x=0 plane.
Once the collision happens,
a dense gas layer forms in the collision mid-plane at x=0.
The layer (not clear if it is a {\it filament} yet),
marginally visible at t=2 Myr,
gradually builds up high-density gas throughout the simulation.
At the same time, the high pressure in the mid-plane expels gas
in opposite directions along the y-axis, as indicated by the 
diverging arrows at y$\sim\pm$10 pc.
After $\sim$4 Myr, sink formation begins. At t=5 Myr, the maximum
sink mass reaches 16.5 M$_\odot$. The sink has the potential of forming
a massive star if the star formation efficiency $\ga50\%$ and only
one star forms. From Figure \ref{fig:zslice} we can also see that 
the sink formation is along the dense layer.
In total, there are 57 sinks at t=5 Myr and 
they concentrate in the middle part of the layer.

\begin{figure*}[htb!]
\centering
\epsscale{1.1}
\plotone{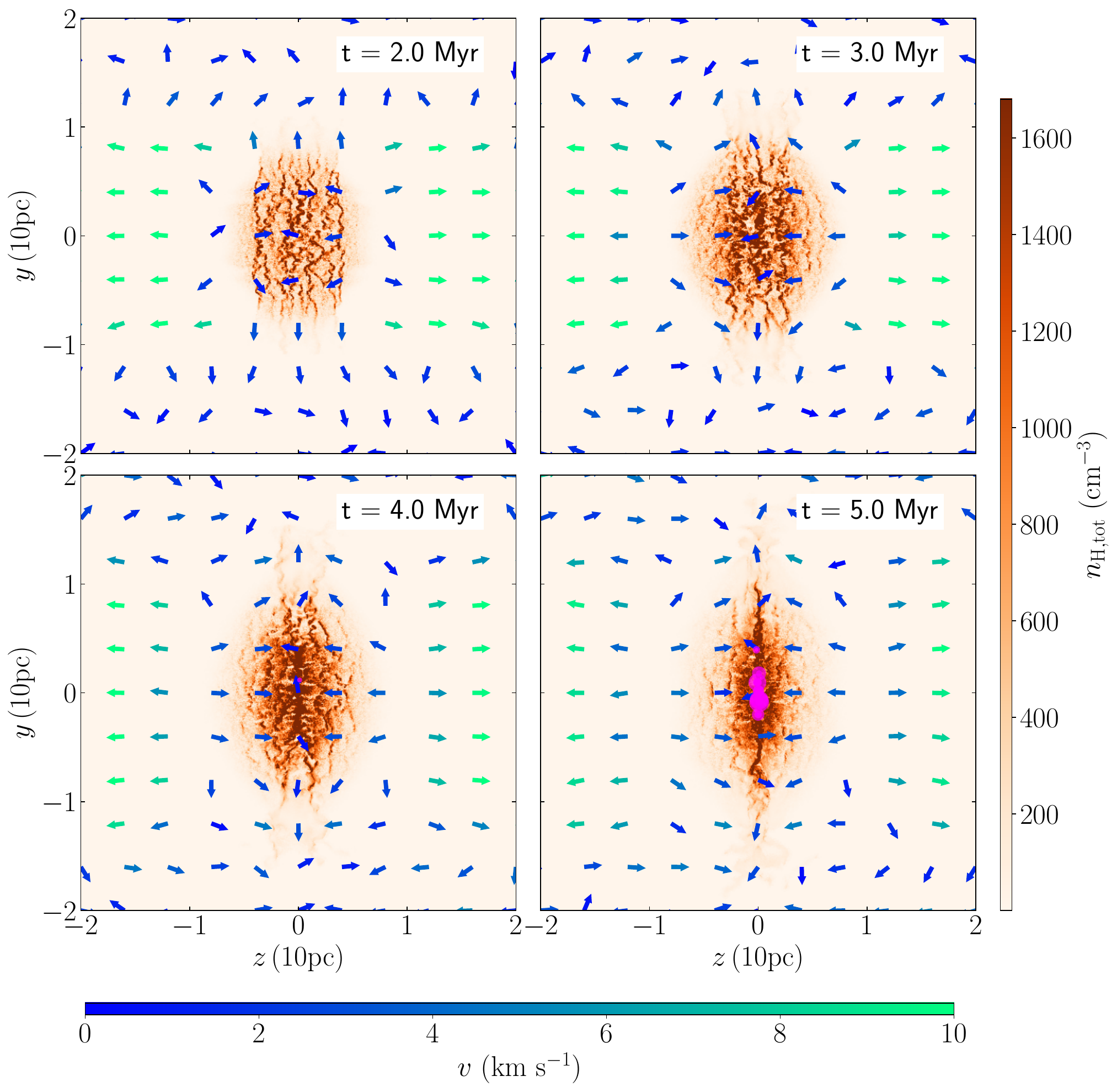}
\caption{
\mbox{MRCOLA\_CNM} density slice plots for the collision mid-plane. 
It corresponds to the x=0 plane in Figure \ref{fig:setup} and
faces the positive x-direction.
The linear color scale shows the total gas density in unit of $n_{\rm H,tot}$.
The magenta filled circles show the sinks. The circle size is proportional to
the sink mass. The time step is shown at the top-right in each panel.
The arrows show the velocity vectors. The arrow color scales with the
velocity magnitude (the bottom color bar).
\label{fig:xslice}}
\end{figure*}

Figure \ref{fig:xslice} shows density slice plots for the x=0 plane.
This is the collision mid-plane that shows up as the projected dense layer
in Figure \ref{fig:zslice}. We will see that the layer indeed becomes
a {\it filament}. In this x=0 plane, the collision produces 
a dense {\it pancake} initially.
The high pressure in this {\it pancake} drives material outward,
which is indicated by the diverging velocity vectors at y$\sim\pm$10 pc. 
Starting from t=2 Myr, converging movements in the {\it pancake} establish,
which is indicated by the velocity vectors toward the central axis.
The approximately symmetric converging movements, 
which are more obvious from t=3 Myr, try to squeeze the {\it pancake}
toward its central axis (the y-axis at z=0), 
while the bipolar outward movements at y$\sim\pm$10 pc remain.
Outside the {\it pancake} at z$\la$-10 pc and z$\ga$10 pc, 
there is a strong outward movement along z-axis as indicated by
the long diverging velocity vectors.
%Material is moving away from the {\it pancake} in z-direction.
Meanwhile, the {\it pancake} interior develops numerous {\it fibers}
roughly parallel to the y-axis (mostly obvious at t=2 Myr).

As shown by K21 (e.g., their Figure 25), the above velocity patterns
are typical results of CMR.
The cloud collision triggers magnetic reconnection
at the outer edges of the {\it pancake} (z$\sim\pm$7 pc). 
The reconnected field forms {\it field-loops} around the {\it pancake}
and half loops for z$<$-7 pc and z$>$7 pc, which is illustrated in 
Figure \ref{fig:setup}(b). The {\it field-loops} pinch the {\it pancake} and make
it a single {\it filament} along the y-axis. The converging velocity arrows in 
Figure \ref{fig:xslice} show the gas movement toward the {\it filament}
due to the pinching. The half loops outside the {\it pancake} pulls gas away, 
resulting in the diverging velocity arrows along z-axis outside the 
{\it pancake}. The locations of the velocity reverse 
(e.g., z$\sim\pm$7 pc at t=1 Myr) are the main CMR sites.

\subsection{Kinematics}\label{subsec:kin}

\begin{figure*}[htb!]
\centering
\epsscale{1.1}
\plotone{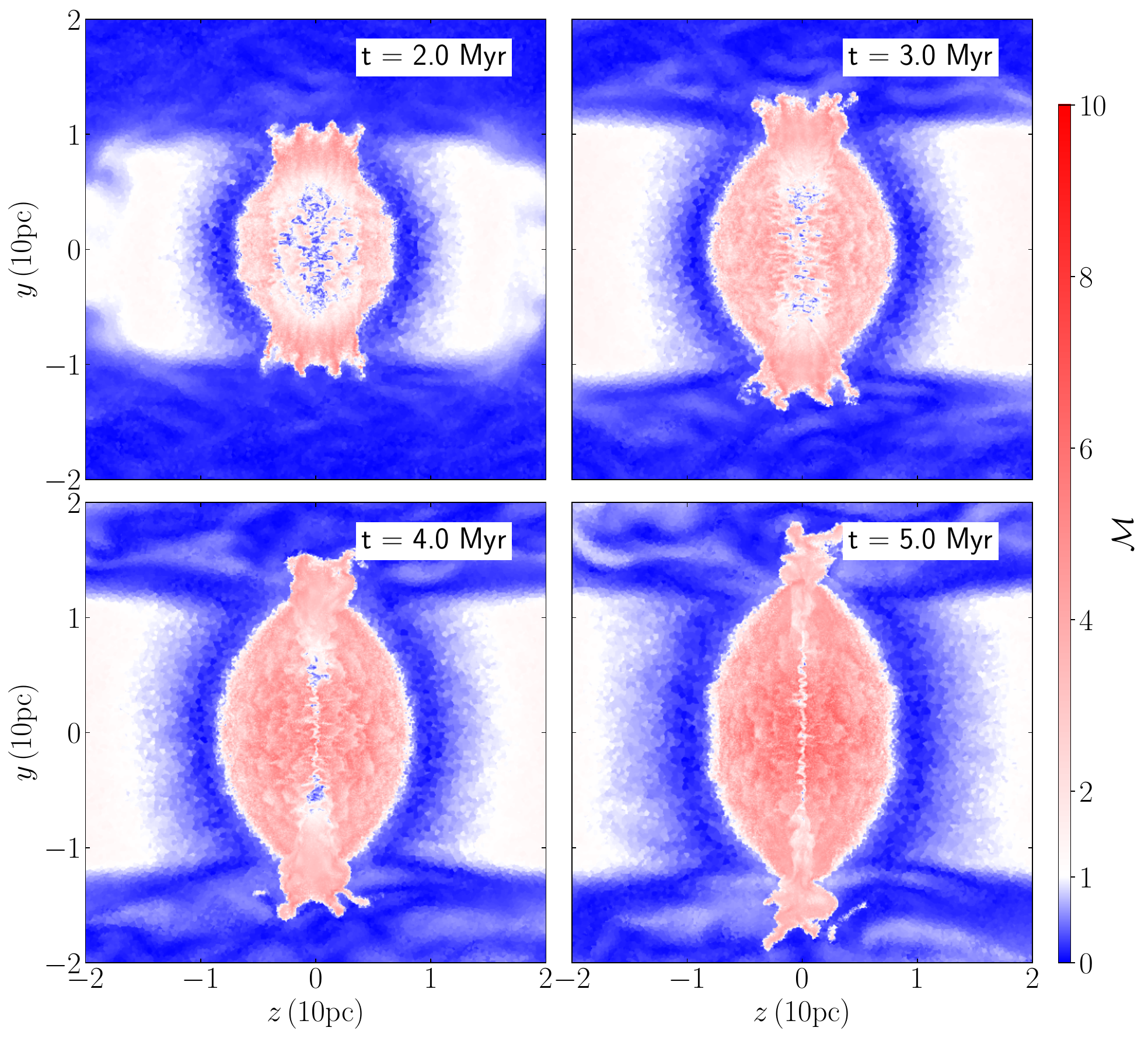}
\caption{
\mbox{MRCOLA\_CNM} sonic Mach number $\mathcal{M}$ for the collision mid-plane (x=0).
This is the same slice as Figure \ref{fig:xslice}.
The diverging color scheme diverges at $\mathcal{M}=1$.
The velocity direction can be depicted from Figure \ref{fig:xslice}.
\label{fig:xsliceM}}
\end{figure*}

We examine the kinematics in the collision mid-plane
and find that the gas velocity is sub-Alfv\'enic
near the reconnection sites
% (Alfv\'en Mach number $\mathcal{M}_A\sim$0.1 at 5 Myr)
but super-Alfv\'enic in the {\it pancake}.
% ($\mathcal{M}_A\sim$20)
The super-Alfv\'enic movement turns out to be supersonic.
Figure \ref{fig:xsliceM} shows the sonic Mach number $\mathcal{M}$
in the mid-plane where $\mathcal{M}\sim1-8$ at t=5 Myr.
Combining with Figure \ref{fig:xslice}, we see that
gas is transported to the central {\it filament} supersonically.
% This is consistent with the velocity vectors shown in Figure
% \ref{fig:xslice} where the sound speed is $\sim$0.4 km s$^{-1}$
% (typical gas temperature 30-50K) in the {\it pancake}.
In the innermost spine of the {\it filament}, the gas slows down
to transonic to moderately supersonic ($\mathcal{M}\sim2$)
because it reaches stagnation
(also see \S\ref{subsec:synobs} and Figure \ref{fig:specsink}).
Note, the temperature in the spine is as low as $\sim$5 K.
Nevertheless, the supersonic transportation in the {\it pancake}
results in supersonic turbulence in and around the {\it filament}.
Note, there is no manually generated turbulence throughout the simulation.
In other words, CMR self-generates supersonic turbulence
as a result of the magnetic transportation of dense gas.
As long as the collision persists, CMR continues to convert colliding
gas into dense gas in the collision mid-plane which is then transported
to the {\it filament}, driving turbulence within the {\it filament}.
CMR thus converts large-scale coherent
motion (the laminar collision) to small-scale turbulence.
%Again, the movie {\tt CMRfield.mpg} 
%\href{https://doi.org/10.7910/DVN/CXHWRR}{here}
%clearly shows the dynamical process.
A dedicated study of the nature of the turbulence will be
carried out in the future, including the scale dependence and
perhaps the power spectrum (in comparison with other models).

Meanwhile, the magnetic reconnection triggers plasmoid instability
\citep[see, for example,][and K21]{2009PhPl...16k2102B}
in each z-x plane at the field-reversal interface. The plasmoids across
multiple z-x planes constitute the {\it fibers} in the {\it pancake}.
With the progress of CMR, the {\it fibers} (also the entire {\it pancake})
gradually converge to the middle symmetric axis
(the y-axis in our coordinate system). At t=5 Myr, the single main 
($\sim$20 pc) {\it filament} forms along the axis. The maximum density 
reaches $n_{\rm H,tot}\sim10^{9}$ cm$^{-3}$, while the typical density
is $n_{\rm H,tot} \ga 2\times10^{3}$ cm$^{-3}$.
At two ends of the {\it filament},
there are {\it forks} that are reminiscent of those structures
at both ends of the Orion A cloud and the California cloud 
\citep[see Figure 6 of][]{2009ApJ...703...52L}. 
Around the main {\it filament}, there is a lower density ``halo'' with 
$n_{\rm H,tot} \la 2\times10^{3}$ cm$^{-3}$ (for t$\ga$ 4 Myr).
This ``halo'' is the relic of the {\it pancake}. There are rich accompanying
{\it fibers} around the ``halo'' and the {\it filament}.
Combining Figures \ref{fig:xslice} and \ref{fig:zslice}, we can see
that \mbox{MRCOLA\_CNM} indeed produces a large filamentary cloud along y-axis.
Star formation happens along this {\it filament}. Note, the length scale of the
{\it filament} is about 10 times larger than that from K22, which is inline with 
the scale increase of the simulation. 

\begin{figure*}[htb!]
\centering
\epsscale{1.1}
\plotone{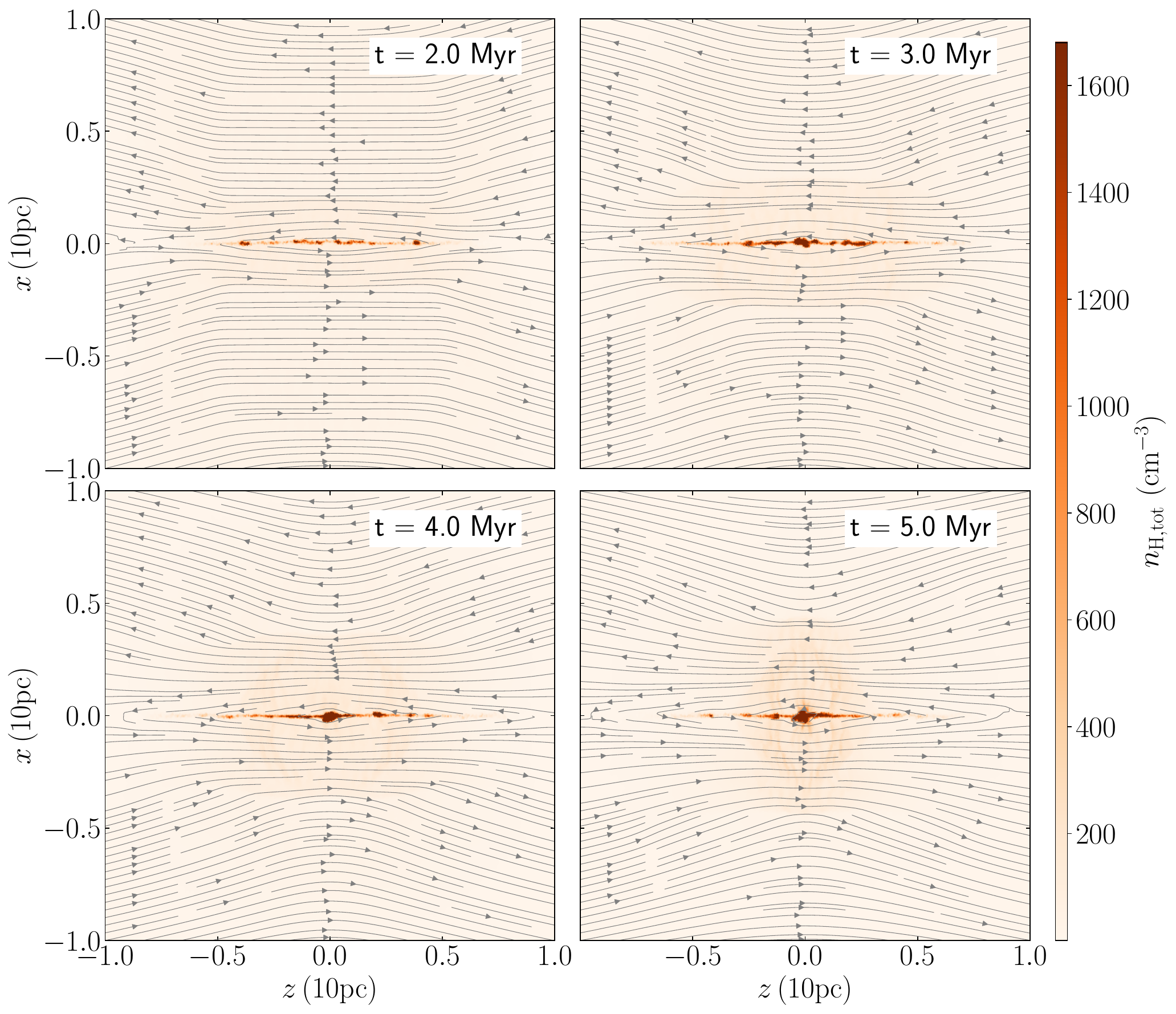}
\caption{
\mbox{MRCOLA\_CNM} density slice plots for the y=0 plane
(compared to Figure \ref{fig:setup}b). 
The linear color scale shows the total gas density in unit of $n_{\rm H,tot}$.
The vector stream lines show the field lines in the plane.
Note how the reconnected lines encircles the high-density {\it pancake} at x=0.
The time step is shown at the top-right.
This slice corresponds to Figure \ref{fig:setup}(b).
\label{fig:yslice}}
\end{figure*}

Figure \ref{fig:yslice} provides a better view of the plasmoids
and the squeezing of the {\it pancake}. Here we show slice plots of y=0 plane,
so we see the cross-section of the {\it pancake} and the {\it filament}.
We can see {\it field-loops} forming around the {\it pancake}
(with sharp turning points at z$\sim\pm7$ pc).
Inside the {\it field-loop}, there are multiple over-dense spots along
the {\it pancake}. The spots are cross-sections
of the {\it fibers} as shown in Figure \ref{fig:xslice}.
The {\it field-loop} formation is due to magnetic reconnection at two ends of
the {\it pancake} and the spots are follow-up over-densities due to the
plasmoid instability at the field-reversal interface. Again,
see K21 for a detailed description (e.g., their Figures 24, 25). 
The reconnected {\it field-loops} continuously squeeze the {\it pancake} to
the central axis. Consequently, from t=3 Myr, we see the accumulation
of gas into the main {\it filament}.
Throughout the collision, {\it field-loops} continuously
transport dense gas to the main {\it filament}. Simultaneously, star formation
happens along the {\it filament} length, showing a dynamic star formation picture.

\subsection{Molecular Gas Formation}\label{subsec:molgas}

\begin{figure*}[htb!]
\centering
\epsscale{1.1}
\plotone{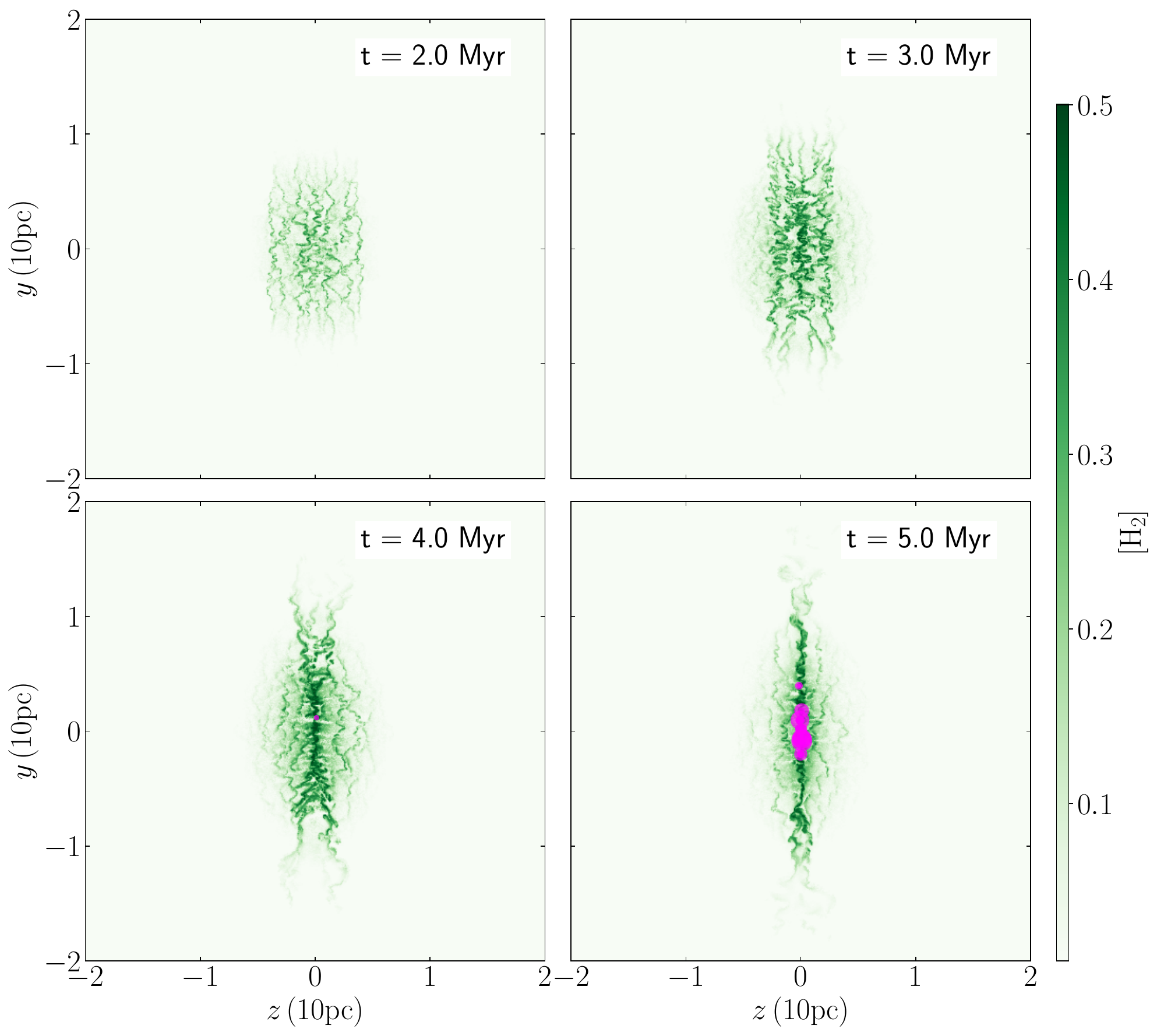}
\caption{
\mbox{MRCOLA\_CNM} molecular fraction [H$_2$] slice plots
for the collision mid-plane (x=0). 
This is the same slice as Figure \ref{fig:xslice}.
The linear color scale shows [H$_2$] from 0 to 0.5 (fully molecular).
The magenta filled circles show the sinks. The circle size is proportional to
the sink mass. The time step is shown at the top-right.
\label{fig:xh2}}
\end{figure*}

A major goal of this paper is to test whether a molecular cloud
can form through CMR. The first quantity we examine is the molecular gas
fraction [H$_2$]$\equiv n_{\rm H_2}/n_{\rm H,tot}$. If all H atoms are in 
molecular form, i.e., the {\it filament} is fully molecular, the molecular
fraction [H$_2$] should be 50\%. If the {\it filament} is fully atomic, [H$_2$] should
be simply 0. From previous sections we have seen 
that CMR gradually creates the {\it filament}.
Here we examine its molecular fraction as a function of time.

Figure \ref{fig:xh2} shows the time-dependent [H$_2$] from our fiducial model
\mbox{MRCOLA\_CNM}. Initially, [H$_2$] is simply zero because the model started
with all H in atomic form. At t=2 Myr, we begin to notice an elevated [H$_2$]
in the central {\it pancake}. We have seen the dense {\it pancake} in 
\S\ref{subsec:density} and Figure \ref{fig:xslice}.
There the {\it pancake} developed {\it fibers}.
Here, similar {\it fibers} are also visible in [H$_2$],
indicating an ongoing molecular formation.
With the progress of the simulation, [H$_2$] elevates in the {\it fibers}
which merge into the central main {\it filament}. The two ends of the main {\it filament}
also show the {\it fork} morphology. Sink formation happens along the {\it filament}
near the central part. At t=5 Myr, a single {\it filament} with [H$_2$] approaching 0.5
(fully molecular) is present with sinks in the middle. 
By this time, the total molecular mass is $\sim$800 M$_\odot$.

While the density plot
showed a ``halo'' around the center of the {\it filament} (Figure \ref{fig:xslice}, 
\S\ref{subsec:density}), the molecular enhancement in the [H$_2$] plot is more
like a pure {\it filament} because the density ``halo'' is not dense enough to
produce a large amount of molecular gas. Regardless, Figure \ref{fig:xh2} shows
that CMR of CNM indeed creates a giant molecular cloud ($\sim$20 pc).

\begin{figure*}[htb!]
\centering
\epsscale{1.1}
\plotone{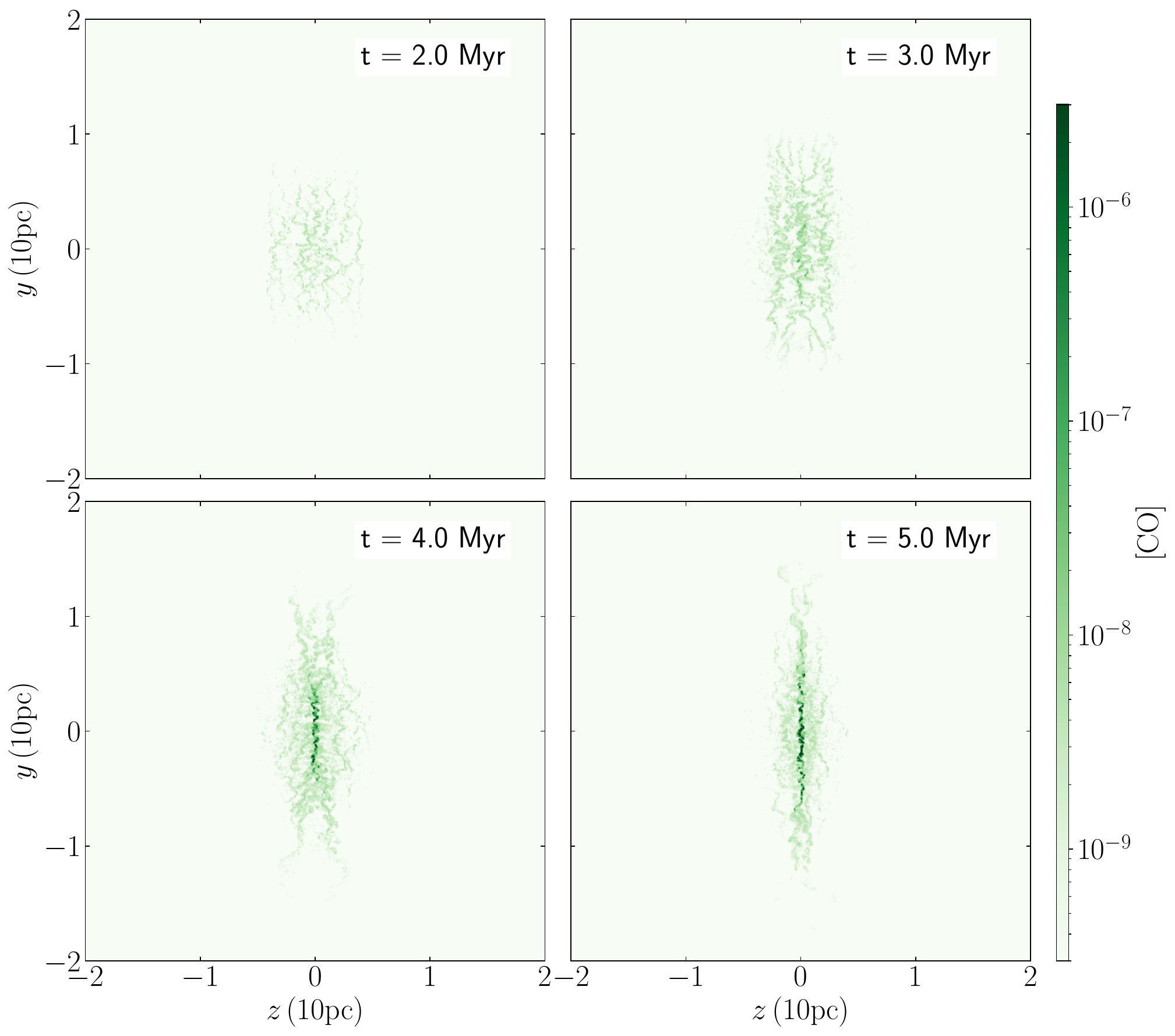}
\caption{
\mbox{MRCOLA\_CNM} molecular abundance [CO] slice plots
for the collision mid-plane (x=0). 
This is the same slice as Figure \ref{fig:xslice}.
The logarithmic color scale shows [CO] from 
$3\times10^{-10}$ to $3\times10^{-6}$.
The time step is shown at the top-right.
\label{fig:xco}}
\end{figure*}

A molecular cloud is typically observed in CO and its isotopologues.
So in Figure \ref{fig:xco} we show the CO abundance
[CO]$\equiv n_{\rm CO}/n_{\rm H,tot}$ development as a function of time.
Our chemical module is based on \citet{Gong17}. Although they used a
simplified chemical network, their [CO] agreed well with the more 
sophisticated PDR code \citep{1985ApJ...291..722T}. 
Our implementation includes several 
additional reactions that make the network more robust when dealing with hot, 
shocked gas. Full details of these modifications can be found in 
\citet{2023MNRAS.519.4152H}. 
The H$_2$ formation in \citet{Gong17} is robust in the sense that
it agrees with \citet{1985ApJ...291..722T,1999ApJ...524..923N}.

In \mbox{MRCOLA\_CNM}, initially all Carbon is in C$^+$ and there is no CO. 
Following \citet{2010MNRAS.404....2G}, we adopt an initial elemental
abundance for carbon as 1.4$\times$10$^{-4}$ and for oxygen as
3.2 $\times$10$^{-4}$ \citep{2000ApJ...528..310S}.
Oxygen is all in atomic form which is subsequently tracked
by conservation laws of abundance and charge \citep{Gong17}. 

As shown in Figure \ref{fig:xco}, 
[CO] is marginally noticeable at t=2 Myr
when the maximum [CO] reaches $\sim6.9\times10^{-8}$ and
the maximum density $n_{\rm H,tot}\sim9.2\times10^3$ cm$^{-3}$.
With the progress of the simulation, dense {\it fibers}
build up and merge into the main {\it filament}. The morphology of
[CO] enhancements follows the density and [H$_2$] enhancements.
The morphological match indicates that H$_2$ and CO formation quickly
follows the accumulation of dense gas because the chemical timescale
is relatively short, which is consistent with the findings in 
\citet{2010MNRAS.404....2G} who used a more sophisticated chemical
network and started with a condition similar to CNM. Finally at t=5 Myr, 
the [CO] enhancement appears as a single {\it filament} without the ``halo'',
simply because the ``halo'' is not dense enough compared to the {\it filament}.
The {\it filament} and its immediate surroundings have [CO]$\ga10^{-8}$,
with innermost clumps having [CO]$\sim1.4\times10^{-4}$, i.e.,
all carbon in CO\footnote{Note that CO depletion is not included.}.
These results confirm the earlier [H$_2$] result that CMR creates a single
filamentary giant molecular cloud with a fully molecular spine.
Again, our findings show that 
a molecular cloud can emerge from a CNM gas collision
between reverse magnetic fields, as long as the condition allows CMR to
produce enough dense gas. 

\subsection{Synthetic observations}\label{subsec:synobs}

\begin{figure*}[htb!]
\centering
\epsscale{1.1}
\plotone{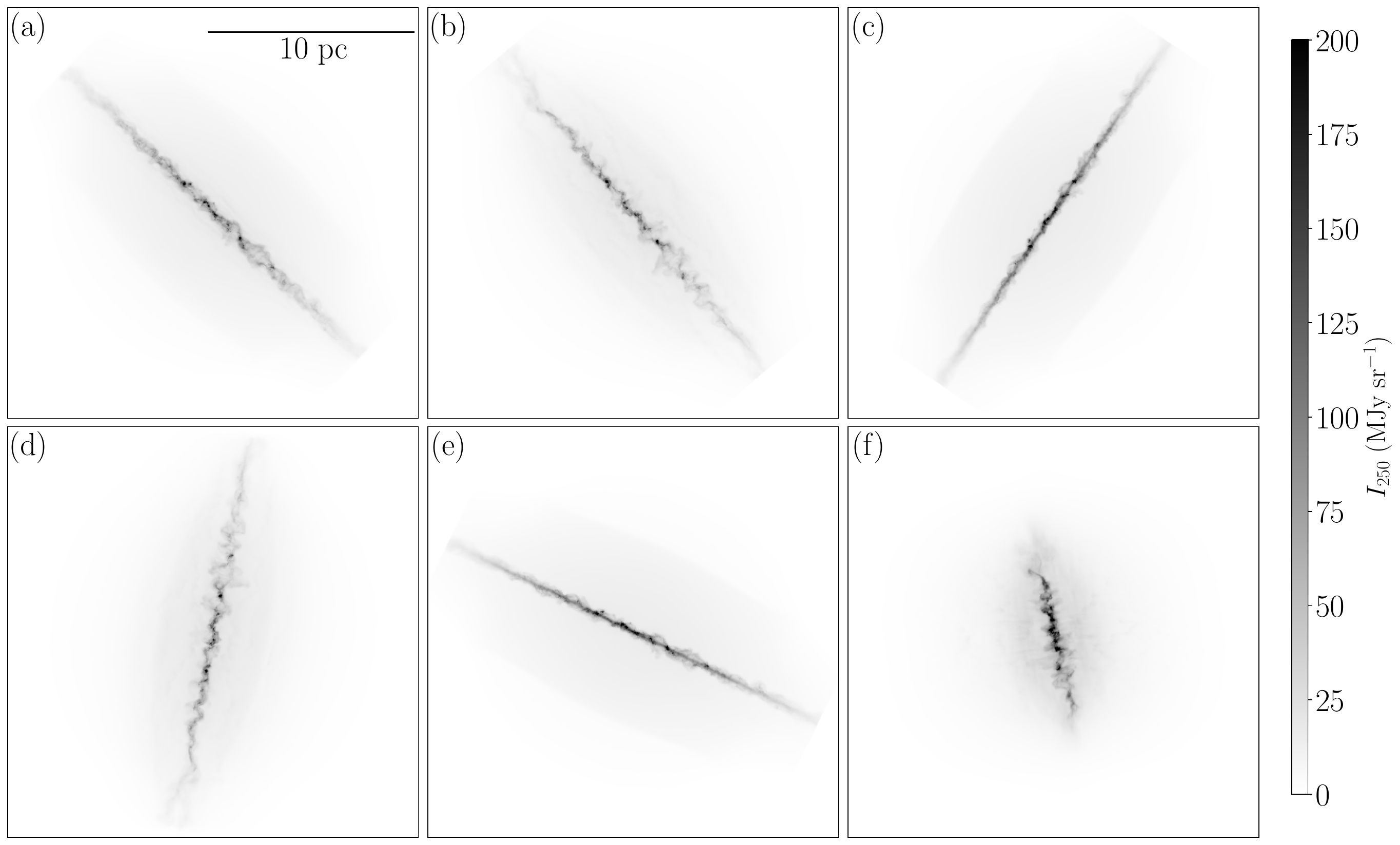}
\caption{
\mbox{MRCOLA\_CNM} synthetic observations of dust emission at 250 $\mu$m at t=5 Myr.
In each panel we view the entire {\it filament} ($\sim$20 pc) at a random orientation
(not a slice plot).
The pixel scale is 0.02 pc, corresponding to 10\arcsec\ at a distance of 400 pc.
The color bar has units of MJy sr$^{-1}$.
\label{fig:cont250}}
\end{figure*}

\subsubsection{Dust emission}

To show the dust emission at far infrared wavelengths, 
we carry out radiative transfer (RT) modeling of the {\it filament}.
We use the RT code RADMC-3D \citep{2012ascl.soft02015D} for this purpose.
The standard dust-to-gas mass ratio of 0.01 is adopted.
The dust is modeled with an amorphous spherical silicate
grain with radius 0.1 $\mu$m. The opacity at 250 $\mu$m is a
factor of 4 smaller than the thin ice mantle model from
\citet{1994A&A...291..943O} in which dust evolution was
considered. So our dust emission, if optically thin, could
be underestimated by the factor if there was grain growth.
In our simulation,
the majority of the dust has a temperature of $\sim16$ K
because it is in equilibrium with the ISRF. In the spine of
the {\it filament}, the dust temperature is as low as $\sim5$ K
due to the high extinction.
Figure \ref{fig:cont250} shows the 250 $\mu$m emission 
for different viewing angles. The pixel scale is 0.02 pc
and no beam convolution is applied. 
Nor do we add any noise.
% The intensity is in unit of MJy sr$^{-1}$.

As shown in Figure \ref{fig:cont250}, the morphology of the dense
gas is indeed filamentary, with a length of 10-20 pc
depending on the projection. In all panels, the {\it filament}
shows {\it fibers}.
Panels (a)(c)(e) show {\it rings} along the {\it filament}.
Panels (b)(d) show a {\it fork} at one end of the {\it filament}.
Multiple dense cores show up as dark dots in the central portion
of the {\it filament}.
% As will be shown in Figure \ref{fig:zoomcont250},
% the dense cores host sinks that are actively accreting.
In general, the plane-of-sky magnetic fields are perpendicular to the filament, while the line-of-sight fields reverse on two sides of the filament, similar to the scenario in Orion A \citep{1997ApJS..111..245H,2019A&A...629A..96S}.

\begin{figure*}[htb!]
\centering
\epsscale{1.1}
\plotone{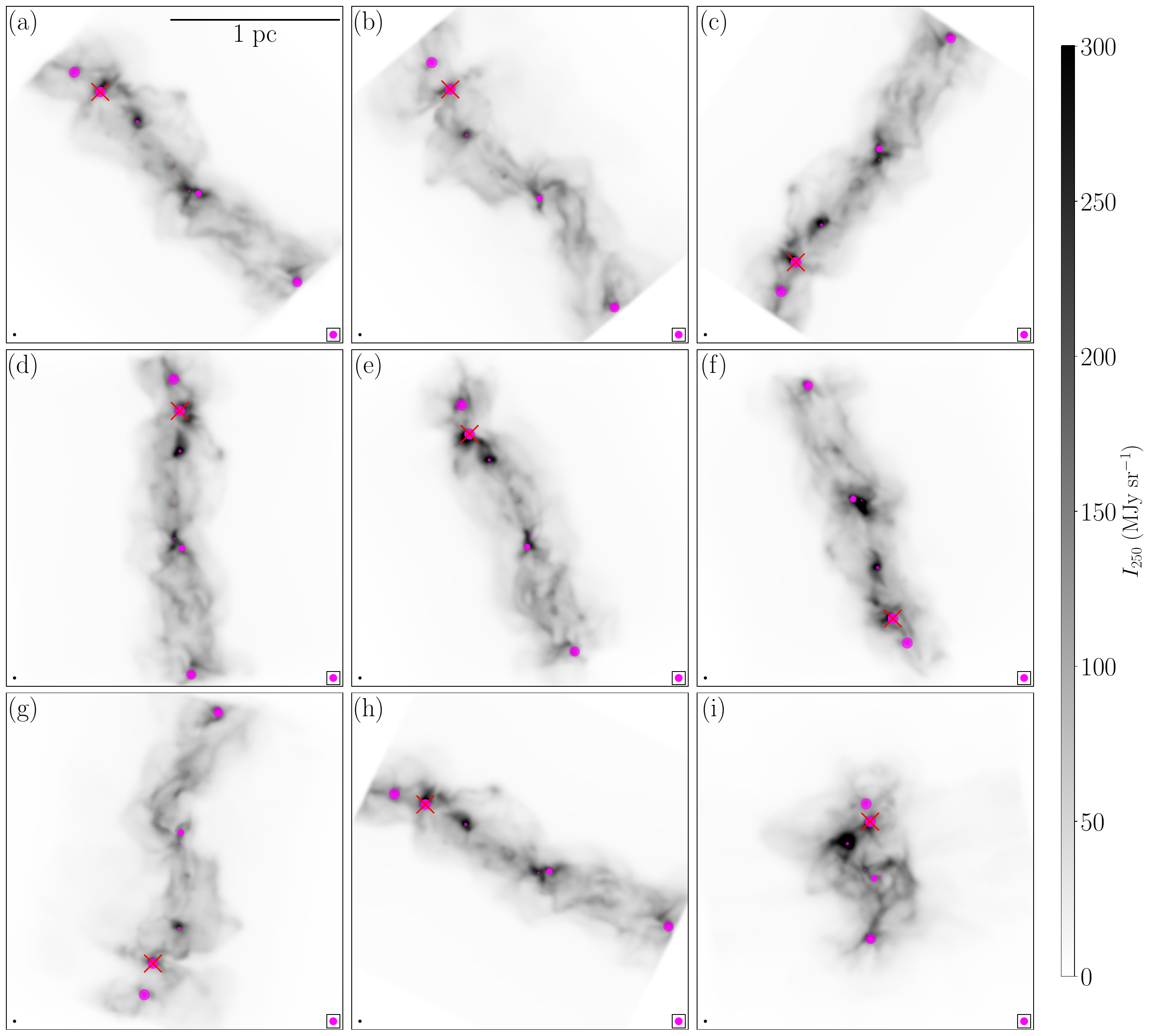}
\caption{
Zoom-in views of the central 2 pc region of the
250 $\mu$m emission at t=5 Myr.
Again each panel shows a random viewing angle of the {\it filament} (not a slice plot).
The images are smoothed to a spatial resolution of 10\arcsec,
corresponding to 0.02 pc at a distance of 400 pc.
The lower-left black filled circle shows the 10\arcsec\ beam.
The color bar has units of MJy sr$^{-1}$.
The magenta filled circles are the sinks.
Their sizes are proportional to the mass.
The red cross marks the location of sink-31.
The lower-right legend represents an 8 M$_\odot$ sink.
\label{fig:zoomcont250}}
\end{figure*}

Figure \ref{fig:zoomcont250} shows a zoom-in view to the central
2 pc region at t=5 Myr. Again we use RADMC-3D \citep{2012ascl.soft02015D}
to model the 250 $\mu$m emission for different projections.
We adopt a smaller pixel size of 0.004 pc, corresponding to 2\arcsec\
at a distance of 400 pc. To match the resolution in Figure \ref{fig:cont250},
we smooth the image to 10\arcsec. We overplot the sinks in the RT images.
Their sizes are proportional to their masses.

As shown in Figure \ref{fig:zoomcont250}, the {\it fibers} are
more prominent in the zoomed view. They show a chaotic organization but
constitutes the overall filamentary morphology. The {\it fibers}
show up in all projections. The overall
{\it filament} is like a bundle of {\it fibers} thanks to the cylindrical symmetry
of the reconnected helical field.
Although we are not reproducing a particular
molecular cloud, the morphology and the intensity
is reminiscent of the Taurus L1495 filament,
which has typical values around $\sim 100$ MJy/sr
\citep{2013A&A...550A..38P}. The converging gas flow
in the {\it pancake} (Figure \ref{fig:xsliceM} and the
\href{https://doi.org/10.7910/DVN/CXHWRR}{movie})
potentially explains the mass accretion in the
sheet-like gas around the filament \citep{2019A&A...623A..16S}.
Note, however, at the small scale
of {\it fibers}, non-ideal MHD effects may be important. For instance,
\citet{2023ApJS..265...58K} showed that higher Ohmic resistivity
resulted in less {\it fibers} and a smoother overall morphology.
It is also not clear if gas and dust reacts to CMR differently
since K21 showed slightly different morphologies between far infrared
dust emission and molecular line emission (C$^{18}$O) 
in the Stick-filament.

At the intersection of the {\it fibers},
gas may become self-gravitating and collapse,
thus we see dense cores develop with connecting {\it fibers} stretching out.
Sinks form in the dense cores, suggesting that they are 
capable of forming new stars. For instance, the sink at the top-left corner
in panel (a) shows a dusty tail. The sink next to it in the
same panel shows multiple connecting {\it fibers}.
In fact, there are five sinks at this location,
with the most massive one (hereafter sink-31) having 15 M$_\odot$
(marked with the red cross). In all other panels, sink-31 shows multiple connecting
dusty {\it fibers}. Its mass ($>$8 M$_\odot$) shows the
potential of massive star formation (depending on the
star formation efficiency which is not modeled here).

\subsubsection{Dust polarization}

\begin{figure*}[htb!]
\centering
\epsscale{0.58}
\plotone{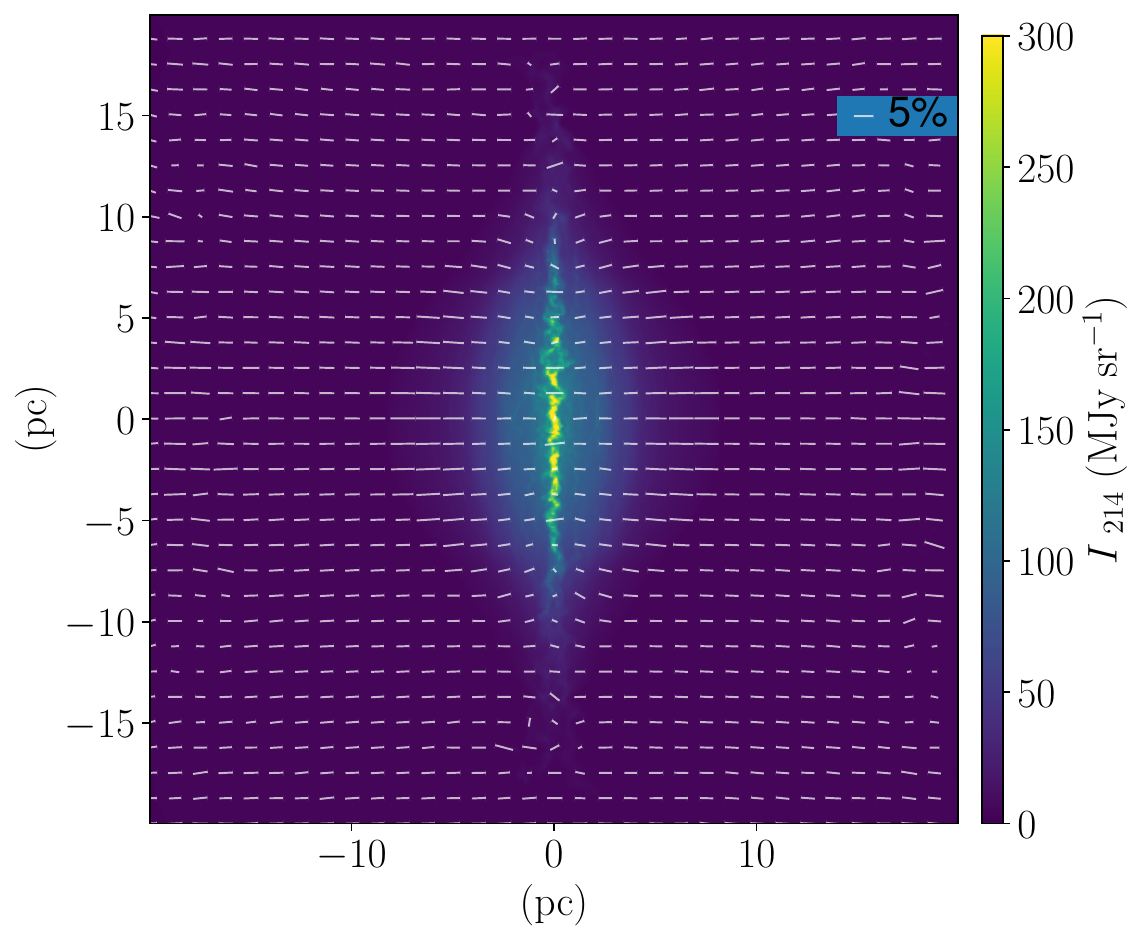}
\epsscale{0.56}
\plotone{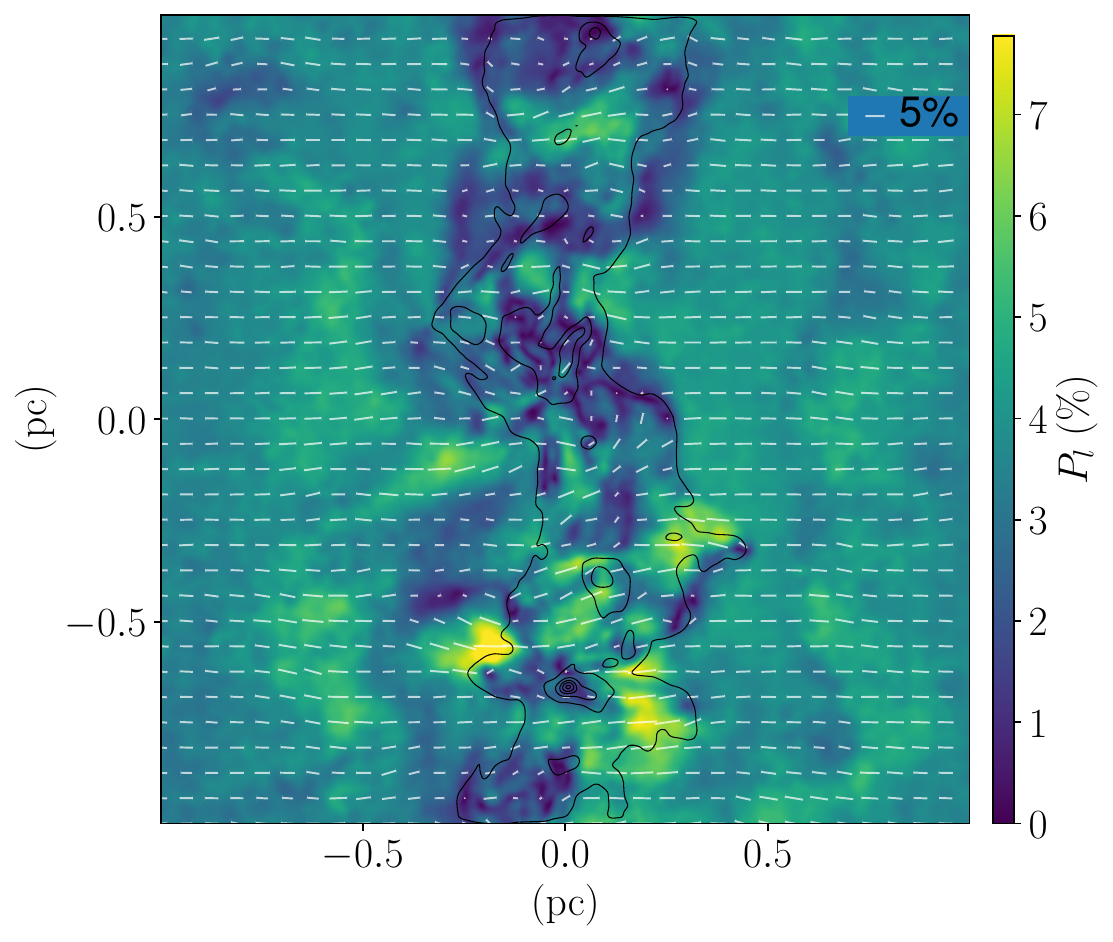}
\caption{
{\bf Left}: Linear polarization $P_l$ overlaid on the stokes I
intensity at 214 $\mu$m. The white segments are oriented along
the plane-of-sky fields. Their lengths are proportional to $P_l$
(5\% $P_l$ labeled at the top-right). The line-of-sight follows
Figure \ref{fig:setup}(a) except that the cube is rotated by
30\arcdeg\ about the y-axis.
{\bf Right}: A zoom-in model of the left panel. The color map
shows $P_l$. The black contours show the stokes I intensity.
The map has a beam size of 10\arcsec.
\label{fig:linpol}}
\end{figure*}

To show the synthetic dust polarization, we utilize the
radiative transfer code POLARIS \citep{2016A&A...593A..87R}.
We adopt a dust model with 62.5\% oblate silicate grains and
37.5\% oblate graphite grains with a power-law size
distribution from 0.005 $\mu$m to 1 $\mu$m
\citep{1977ApJ...217..425M}. The dust is
aligned to magnetic fields via the radiative torque (RAT) 
mechanism \citep{1997ApJ...480..633D,2007MNRAS.378..910L}.
Note, the dust model with POLARIS is different from what
we adopt for the RADMC-3D model. Polarization would be
enhanced if we were to increase the fraction of silicate
grains. Since we are showing general features of
CMR-filaments and not reproducing specific clouds, we
keep both dust models without resolving the inconsistency.

Figure \ref{fig:linpol} shows the dust polarization at 214 $\mu$m
from the POLARIS modeling. We first show the overall field
orientation relative to the {\it filament} in the left panel.
One can see that the field is perpendicular to the {\it filament}.
This result is not surprising as the initial field symmetry
determines the {\it filament} formation, i.e., a CMR-filament
must form along the direction that is perpendicular to the initial
antiparallel field (see Figure \ref{fig:setup}). Observationally,
this implies that we shall see the plane-of-sky field
perpendicular to a CMR-filament if the latter is perpendicular
to our line-of-sight. This scenario is consistent with the 
observations in the Orion A integral-shaped filament
\citep[ISF,][]{1987ApJ...312L..45B,2019A&A...629A..96S}. However,
if a CMR-filament is inclined, the relative orientation in the
plane-of-sky can be oblique simply because of a geometry effect.
If two vectors are perpendicular in 3D, their projected relative
orientation in 2D can be any angle between 0 and 180\arcdeg\
\citep[e.g., see][]{2017ApJ...846...16S,2019ApJ...874..104K}. 
Interestingly, the part of Orion A to the south of the
ISF shows such an oblique filament-field
orientation \citep{2019A&A...629A..96S}.

In the right panel, we zoom in to the central 2 pc region.
Again, we see that the {\it filament} is perpendicular to
magnetic fields. Two new features are worth mentioning.
First, the field orientation in the {\it filament} is disrupted.
This feature is manifested by the randomized segments inside the
{\it filament}. The disruption is a result of the chaotic nature
of the reconnected field as well as the turbulent gas motion
due to the magnetic transportation (\S\ref{subsec:kin} and
Figure \ref{fig:xsliceM}). In line with this physical process,
the polarization fraction $P_l$ becomes highly fluctuated,
which is the second feature as shown in the color map. In
contrast, models without CMR typically show smoother $P_l$
\citep[for example, see][]{2021MNRAS.500..153R}. Therefore, one
might distinguish CMR from other mechanisms by characterizing
their spatial fluctuation frequencies. We defer this analysis
to a future study.

We caution here that the dust polarization in and around a
forming molecular cloud is subject to a number of factors,
in particular the dust composition and evolution (which is
not tracked in our simulation). Observationally, polarization
detection is also limited by several factors (e.g., sensitivity,
wavelength). So the above polarization synthetic observation
is only illustrative. A rigorous comparison of polarization
between different models may be useful to observationally
distinguish them, which is out of the scope of this paper.

\begin{figure*}[htb!]
\centering
\epsscale{1.18}
\plotone{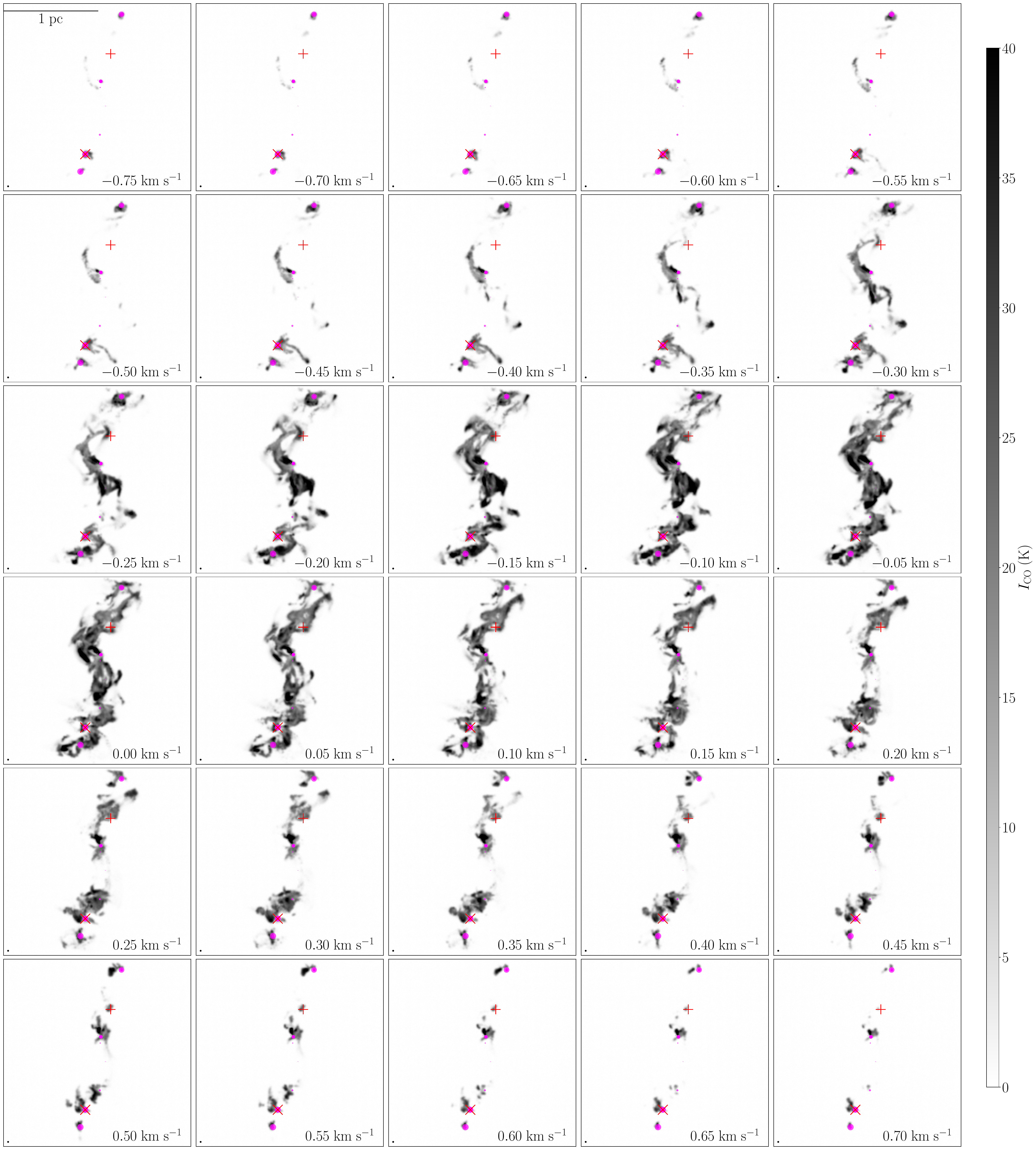}
\caption{
CO(1-0) channel maps from RT modeling of Figure \ref{fig:zoomcont250}(g).
The color scale shows the emission intensity in K. 
To match Figure \ref{fig:zoomcont250}, the channel maps
are smoothed to a Gaussian beam with a FWHM of 10\arcsec,
which is shown at the lower-left corner in each panel.
The channel width is 0.05 km s$^{-1}$, and the channel velocity
is shown at the lower-right corner. The magenta filled circles
are the same as those in Figure \ref{fig:zoomcont250}(g).
The red cross marks the location of sink-31.
The red plus marks the location of no-sink-1.
\label{fig:zoomch}}
\end{figure*}

\subsubsection{Line emission and kinematics}

To show the gas component and the kinematics of the {\it filament},
we carry out RT modeling of CO(1-0) and show the channel maps
in Figure \ref{fig:zoomch}. Again, we use the RADMC-3D code.
Here, we use the non-LTE method with ``linemode=3''
(see the RADMC-3D \href{https://www.ita.uni-heidelberg.de/~dullemond/software/radmc-3d/manual_radmc3d/index.html#}{documentation} for details).
We set H$_2$ as the collision
partner without distinguishing between ortho- and para-H$_2$.
Depending on the density, the temperature, and the linewidth, a CO 
column density of 10$^{17}$ cm$^{-2}$ is marginally optically thick in CO(1-0),
with an optical depth $\tau\sim2$.
In the simulation domain, more than 99\% of the area has a column density
below 10$^{17}$ cm$^{-2}$. Therefore, the majority of the domain is 
still optically thin, except for several dense clumps in the {\it filament}. 
So the channel maps should show the 
true internal structures of the {\it filament} for the most part.
The highest CO column density reaches $3.3\times10^{20}$ cm$^{-2}$.
Modern millimeter telescopes with
advanced spectrometers can typically reach a spectral resolution
of 0.05 km s$^{-1}$, which we adopt for our line cube. 
We also match the spatial resolution to Figure \ref{fig:zoomcont250}.

\begin{figure}[htb!]
\centering
\epsscale{1.18}
\plotone{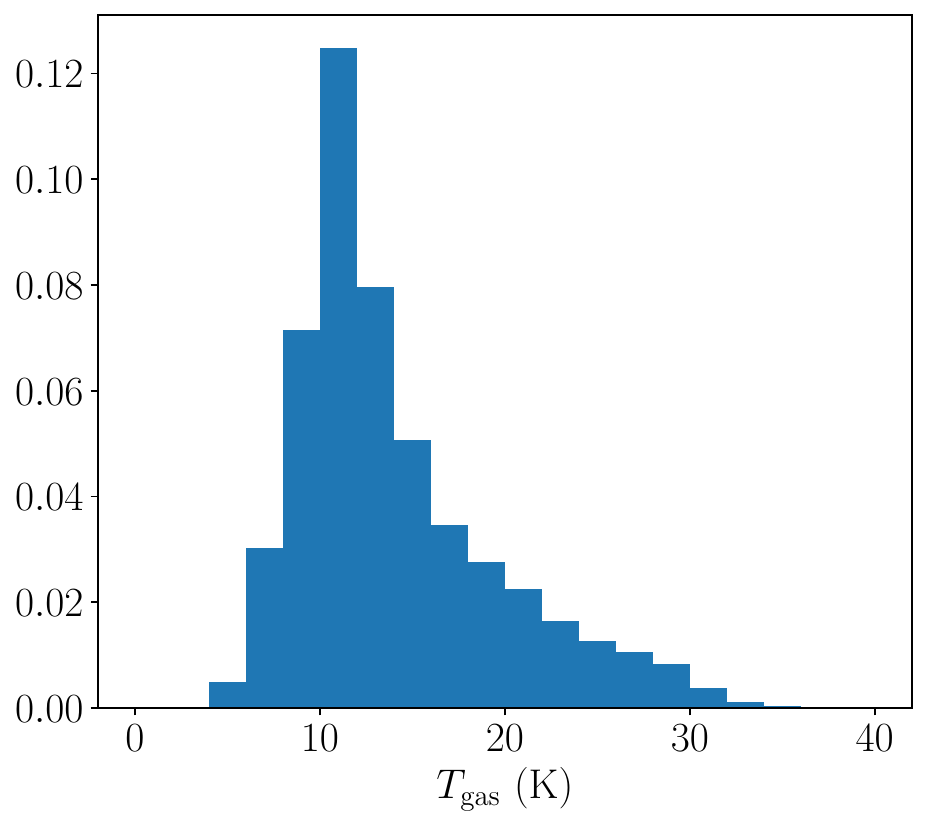}
\caption{
CO-dominated gas temperature histogram in K.
The CO-domination is defined as CO abundance greater than
half the initial elemental abundance of Carbon.
\label{fig:Tf}}
\end{figure}

As shown in Figure \ref{fig:zoomch}, for the majority of the CO gas,
CO(1-0) emission shows a span of about 1 km s$^{-1}$, except for
gas around sinks which shows higher velocities.
Since we set up the fiducial model
\mbox{MRCOLA\_CNM} as a head-on collision, the {\it filament} is at
a stagnation point, thus the limited velocity range in line emission.
An off-axis collision (e.g., K21 and K22) should result in a rotation 
of the {\it filament}, thus a wider velocity span. 
Still, there are multiple velocity components in the channel maps.
Here, the general morphology of the CO emission appears to be turbulent
with rich sub-structures, mostly {\it fibers} and clumps.
Several {\it fibers} and clumps are clearly coherent, e.g., 
the {\it fiber} in the middle of the first row and the two {\it fibers}
in the middle and the bottom of the second row.
The rich sub-structures over multiple velocities are consistent with
the aforementioned supersonic turbulence.

\begin{figure}[htb!]
\centering
\epsscale{1.1}
\plotone{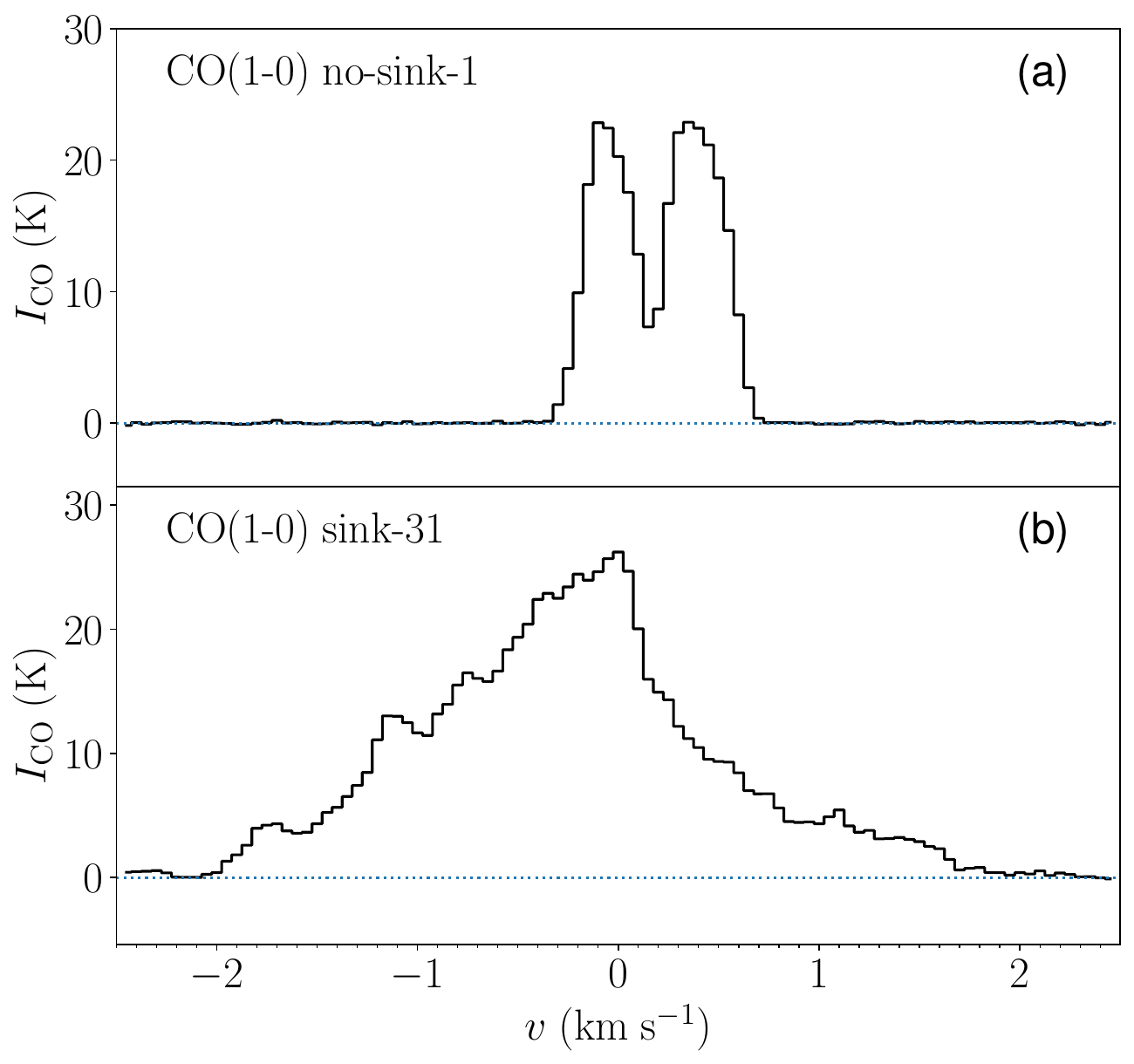}
\caption{
CO(1-0) spectra at two locations (with and without a sink)
in Figure \ref{fig:zoomch}. The spectral cube is from the RT in 
\S\ref{subsec:synobs}.
{\bf (a):} CO(1-0) spectrum within a 10\arcsec\ beam centered at no-sink-1.
{\bf (b):} CO(1-0) spectrum within a 10\arcsec\ beam centered at sink-31.
\label{fig:specsink}}
\end{figure}

In Figure \ref{fig:specsink}(a), we show the CO(1-0) spectrum
at a location without sinks. The location is marked as the plus sign
in Figure \ref{fig:zoomch}. The spectrum shows double velocity
components with a velocity separation $\sim$1 km s$^{-1}$. To compare
this with the sound speed, we plot a histogram of CO-dominated gas
temperatures in Figure \ref{fig:Tf}. The CO-domination is defined as
CO abundance greater than half the initial elemental abundance of Carbon
(see \S\ref{subsec:molgas}). We can see the CO-dominated gas has a
temperature ranging from 5 K to 35 K, with a peak around 11 K. The
peak temperature corresponds to a sound speed of $\sim$0.2 km s$^{-1}$.
Therefore, the CO-dominated molecular gas shows supersonic turbulence.
It is also consistent with what we found in \S\ref{subsec:kin} and
Figure \ref{fig:zoomch}.

% As we have mentioned in \S\ref{subsec:kin}
% and Figure \ref{fig:xsliceM}, dense gas is
% transported to the {\it filament} supersonically in the {\it pancake}.
% So the {\it filament} must have supersonic turbulence due to the
% transportation. The magnetic transportation continues as the collision
% continues, providing a constant source of turbulence to the {\it filament}.
% Dense sub-structures in and around the {\it filament} likely originate
% from shocks. The shocked dense gas is transported to the
% central {\it filament} piece by piece.
% See the movie {\tt CMRfield.mpg} 
% \href{https://doi.org/10.7910/DVN/CXHWRR}{here}
% for an illustration of the transportation.
% Therefore, at least in \mbox{MRCOLA\_CNM},
% the CMR-filament naturally has supersonic turbulence
% which has a magnetic origin due to CMR transportation.

The sinks form in dense gas and
some of them are associated with high velocity gas,
which is indicative of the active accretion onto 
the sink particle. For example, sink-31
in Figure \ref{fig:zoomch} (with the dusty {\it fibers}
as in Figure \ref{fig:zoomcont250})
consistently shows associating gas across all channels,
particularly in high-velocity channels.
Figure \ref{fig:specsink}(b) shows the CO(1-0) spectrum
averaged over a 10\arcsec\ beam at sink-31.
It shows a non-Gaussian profile with high-velocity tails.
The highest velocity reaches 2 km s$^{-1}$.
With a mass of 15 M$_\odot$, sink-31 has an escape velocity
of 5.3 km s$^{-1}$ at the sink creation radius of 0.005 pc.
Therefore, the sink is actively accreting material.
Note, the gas temperature around sink-31 is $\sim20$ K,
corresponding to a sound speed of $\sim0.3$ km s$^{-1}$.

Combining the above analyses,
the overall velocity distribution in the {\it filament} is
a combination of the CMR-induced kinematics and local gravitational infall.
Notably, the sink accretion is through streamers. As shown in K22,
the streaming accretion is due to the helical field that hinders spherical
gas inflow, which limits the accretion rate (and thus the star formation rate),
as compared to the same cloud collision but without CMR.

\subsection{Models without CMR}\label{subsec:nocmr}

\begin{figure*}[htb!]
\centering
\epsscale{1.1}
\plotone{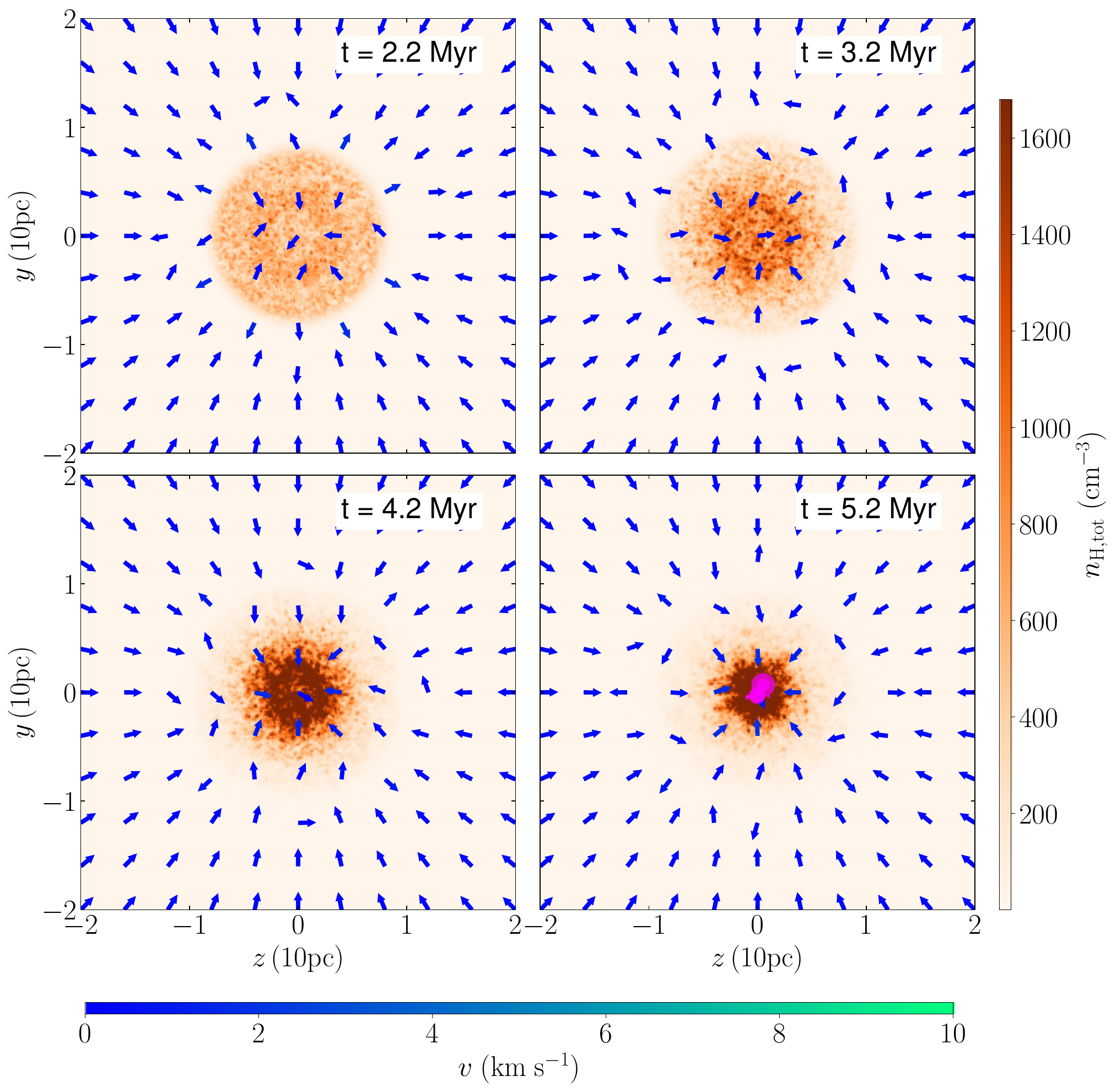}
\caption{
Same format as Figure \ref{fig:xslice} but for model \mbox{MRCOLA\_CNM\_noB}.
\label{fig:xslice_noB}}
\end{figure*}

Similar to K22, we run two control models with the same cloud collision
but without inducing CMR. First, we set up model \mbox{MRCOLA\_CNM\_sameB} in which
the initial magnetic field is uniform. The field strength is the same as
\mbox{MRCOLA\_CNM} but the field vector points toward positive-z throughout
the computational domain. Everything else is the same as \mbox{MRCOLA\_CNM}.
Table \ref{tab:ic} column (3) lists the initial conditions for 
\mbox{MRCOLA\_CNM\_sameB}.
We run the model up to 11 Myr and see no sink formation. 
The maximum density briefly reaches $n_{\rm H,tot}=10^4$ cm$^{-3}$
at 6.4 Myr but ends up reducing to $\sim10^3$ cm$^{-3}$ at 11 Myr.
The lack of dense gas is not unexpected because the dimensionless
mass-to-flux ratio $\mu\equiv(\overline{\Sigma}/B)(2\pi\sqrt{G})$
\citep{1966MNRAS.132..359S,1978PASJ...30..671N}
for the CNM cloud is $\sim$0.85, i.e., magnetically subcritical.
Under such magnetically dominated conditions, dense gas formation is suppressed,
and so is star formation. For instance, in a study of cloud-cloud 
collision, \citet{2020ApJ...891..168W} have shown that star formation
is largely suppressed when the cloud transitions from supercritical to 
subcritical, which is consistent with our results.

Next, we set up \mbox{MRCOLA\_CNM\_noB} in which we model the 
cloud collision in purely hydrodynamics. Table \ref{tab:ic} column (4)
lists the initial conditions for \mbox{MRCOLA\_CNM\_noB}. Besides a zero
initial magnetic field strength, we keep everything else the same as \mbox{MRCOLA\_CNM}.
Figure \ref{fig:xslice_noB} shows the slice plots for x=0 plane from 
\mbox{MRCOLA\_CNM\_noB}. 

Comparing to \mbox{MRCOLA\_CNM} (Figure \ref{fig:xslice}),
the collision in \mbox{MRCOLA\_CNM\_noB} creates a {\it pancake} which does
not become a {\it filament} throughout the simulation. Starting from 
t$\sim$3.2 Myr, the {\it pancake} collapses symmetrically toward the center,
as shown by the velocity vectors. This behavior 
is expected as the head-on collision has symmetry about the collision
axis (x-axis). However, \mbox{MRCOLA\_CNM\_noB} does not have 
the cylindrical symmetry along y-axis as \mbox{MRCOLA\_CNM} 
(determined by magnetic fields) therefore there is
no {\it filament} formation. 
The collapse of the {\it pancake} gives rise to denser gas
at the center of the {\it pancake}, triggering sink formation at 5.2 Myr 
(Figure \ref{fig:xslice_noB} last panel).
The sink formation begins somewhat later than \mbox{MRCOLA\_CNM} (4 Myr).

The difference shows that the turbulent behavior of CMR produces significant
density fluctuations which allows for early dense gas and star formation.
On the contrary, star formation in \mbox{MRCOLA\_CNM\_noB} has to wait
for density accumulation through a gradual, ordered collapse.
Interestingly, K22 found the opposite at a smaller scale and a higher 
initial gas density. They showed that the collision between denser 
molecular clumps formed stars early in the model without magnetic fields.
The transition from CNM to molecular gas probably plays a role,
which shall be
further investigated in detail in the future.
In addition, K22 did not model a head-on
collision. Nevertheless, neither does \mbox{MRCOLA\_CNM\_noB} create
the {\it fibers} in the {\it pancake} (Figure \ref{fig:xslice}), nor
does it form a main {\it filament} as \mbox{MRCOLA\_CNM}.
%, emphasizing the crucial role of CMR in the filament formation process.

\section{Discussion}\label{sec:discus}

\subsection{Comparison to K22}\label{subsec:compareK22}

In this sub-section, we compare our fiducial model \mbox{MRCOLA\_CNM}
with the fiducial model of K22 \mbox{MRCOLA}.
K22 investigated CMR at a factor of 10 smaller scale.
At $\sim$2 Myr, their {\it filament} collapsed into a dense core with a 
cluster of sinks that kept accreting through streamers.
In our case, the sinks also develop the streaming accretion.
However, the gas dynamics in our case are different from K22.
There is no obvious sign of a global collapse in the {\it filament}
up to the ending of the simulation at 5 Myr. The sinks form
at local dense cores that spread out over the main {\it filament}
(Figure \ref{fig:zoomcont250}). On the contrary, the K22 sink
formation was due to the longitudinal collapse of the {\it filament}.
The collapse resulted in a dense core in the middle of the {\it filament}
and sinks formed in the dense core (see K22 Figure 2). Their sinks 
concentrated in the small volume of the dense core ($\sim$0.1 pc, 
see K22 Figure 7), unlike in our case the sinks spread over $\sim$5 pc. 

K22 modeled CMR with an order of magnitude higher initial 
gas density with a fully molecular chemical condition.
So it is understandable that gravity dominated in the K22 
{\it filament} faster than our case. As shown in K22 Figure 2,
their {\it filament} also developed sub-structures with local density
enhancements. It was possible that sink formation could have 
happened in those enhancements if their {\it filament} did not collapse. 
In our model, the global {\it filament} collapse does not have a chance
to develop yet while a few of the sinks are already $>$8 M$_\odot$.
% So massive star formation may happen and counteract the collapse.
Thanks to the complex dynamics induced by CMR, dense cores are able to
form at the intersection of dense {\it fibers}, allowing the formation
of stars without waiting for hierarchical fragmentation to kick in. 
In fact, due to CMR,
the {\it filament} is never a smooth cylinder even at its formation.
So the embedded star formation is not a step-by-step process,
i.e., first reaching an equilibrium before developing instabilities
which lead to local collapse and star formation. At least with 
the one model in this paper (\mbox{MRCOLA\_CNM}), the whole 
process of {\it filament} and star formation is dynamic.

\subsection{Timescale of the molecular cloud}\label{subsec:time}

Sink formation in \mbox{MRCOLA\_CNM} happens around 4 Myr after the
CNM cloud collision. However, the {\it filament} spine is primarily 
molecular by the 4 Myr mark.
The molecular gas fraction is high well before
the sink forms (e.g., Figure \ref{fig:xh2}).
On the other hand, even at 5 Myr, there are only 57 sinks in 
\mbox{MRCOLA\_CNM}, 5 of which with a mass $>$8 M$_\odot$.
However, the Orion A giant molecular cloud has thousands of
young stars (if not more) with a typical age of 1-2 Myr
\citep[e.g.,][]{1997AJ....113.1733H,2021ApJ...908...49F}.
If Orion A formed via CMR as we suspected, then there could be a long,
quiescent phase of the molecular cloud prior to the active star formation.
Alternatively, the collision could have been between higher mass clouds that were already at least partially molecular though maybe CO-dark.
Note, the timescale under discussion is the cloud age before major star 
formation, so the cluster age is not relevant. 

In fact, the freefall timescale of a typical CNM spherical cloud
with $n_{\rm H,tot}\sim30$ cm$^{-3}$ is $\sim8$ Myr. The contraction of
a self-gravitating cloud is slow initially but accelerates later.
Depending on the definition of the starting point, the typical starless
phase of a contracting cloud should be a few Myr. For example,
\citet{2019MNRAS.490.3061V} showed that the 
time interval between the global contraction of CNM and the initial 
low-mass star formation was several Myr. 
Alternatively, if we consider the conversion of HI to H$_2$, the timescale
is $\tau_{\rm HI\rightarrow H_2}\sim10^3/n_{\rm HI}$ Myr, 
assuming an H$_2$ formation rate 
coefficient of $\sim10^{-17}$ cm$^3$ s$^{-1}$. Again, adopting
our CNM density of $n_{\rm H,tot}\sim30$ cm$^{-3}$, the H$_2$ formation
timescale is $\sim30$ Myr. Of course, as the cloud collision and CMR
progress, the density of the {\it filament} increases with time. In our 
fiducial model, the timescale for the starless phase is about 4 Myr.
Observationally, the HI-to-H$_2$ conversion time has been used to estimate
the age of molecular clouds. For instance, \citet{2005ApJ...622..938G}
studied the HI narrow self-absorption \citep[HINSA,][]{2003ApJ...585..823L}
in several molecular clouds
and found that the clouds had ages of 3-10 Myr (or even longer). Their 
assumption of the physical model was that a sudden density enhancement
increased the extinction and greatly reduced photodissociation of H$_2$.
The enhancement triggered the atomic to molecular transition thus the
formation of the molecular cloud. This physical picture is consistent
with a cloud-cloud collision scenario, including the CMR cloud formation
model in this paper. Following the same methodology,
\citet{2020RAA....20...77T} surveyed HINSA in more clouds and found 
similar timescales. 

\begin{figure*}[htb!]
\centering
\epsscale{1.1}
\plotone{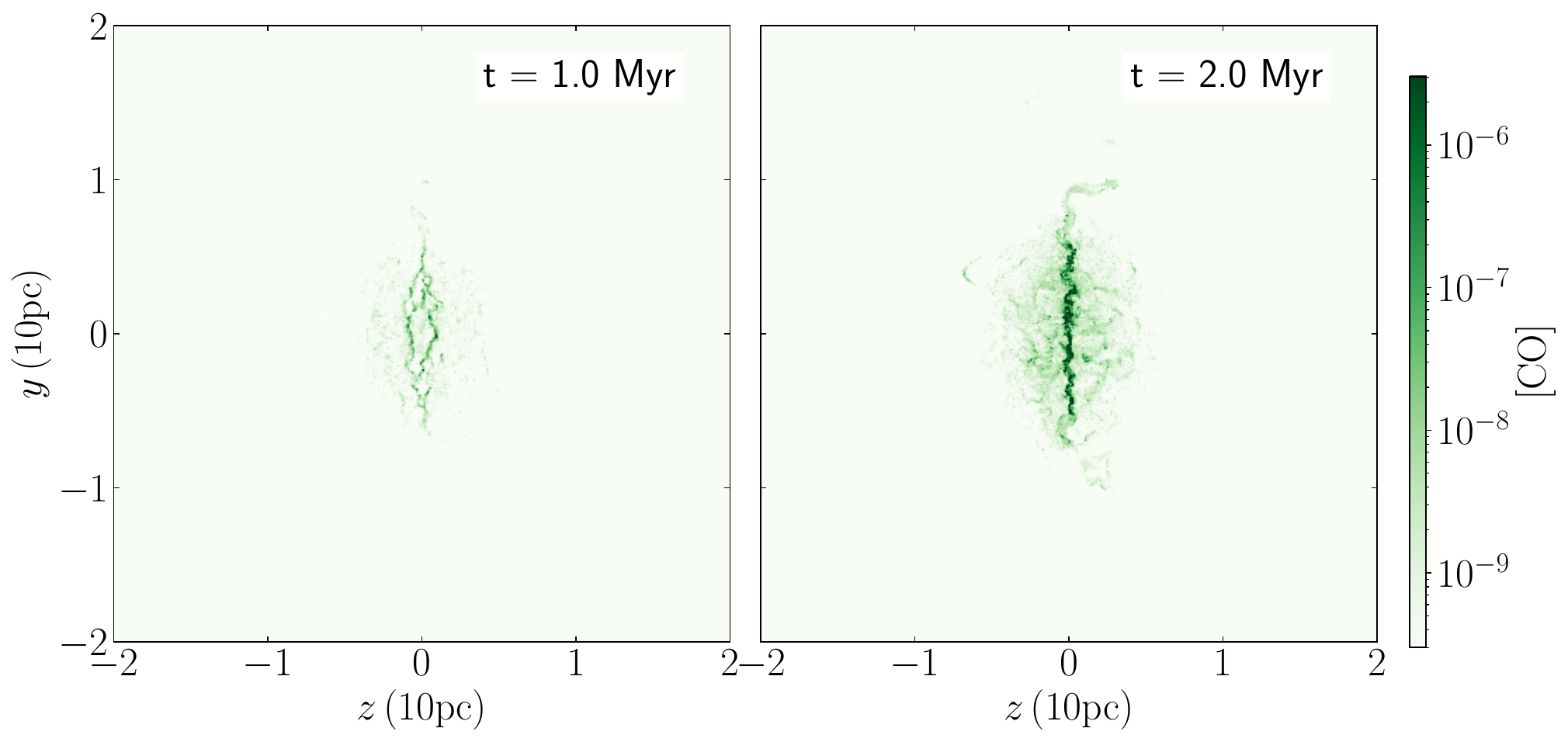}
\caption{
Same format as Figure \ref{fig:xco} but for model \mbox{MRCOLA\_CNM4}
(see \S\ref{subsec:time}).
\label{fig:xco4}}
\end{figure*}

Given the above consideration, the long starless phase of the CMR-filament
in \mbox{MRCOLA\_CNM} is not unexpected. In comparison, the model without
CMR has an even longer starless phase of 5 Myr (\mbox{MRCOLA\_CNM\_noB},
Figure \ref{fig:xslice_noB}). Such a long starless phase is also seen in
the colliding flow simulations of \citet{2012MNRAS.424.2599C}. Their slow
collision model, which is more relevant to \mbox{MRCOLA\_CNM\_noB} (except
that \mbox{MRCOLA\_CNM\_noB} has no seeding turbulence), has a $\sim$10 Myr
time interval between the cold gas formation and star formation.
This phase of the cloud corresponds to the dark molecular gas phase
\citep[e.g.,][]{2010ApJ...716.1191W}, which is also present in the
CMR-filament.

However, if the simulation begins with a higher initial density,
the timescale is much shortened. To show this effect, we run a new
simulation with the initial cloud density and magnetic fields
a factor of 4 higher than the fiducial model, i.e., $n_1=n_2=120$ cm$^{-3}$
and $B_{\rm 1,z}=-B_{\rm 2,z}$ = 20 $\mu$G. These conditions are similar to
those adopted in \citet{2020ApJ...891..168W}. We terminate the simulation
at t=2 Myr. Hereafter, we name this new model \mbox{MRCOLA\_CNM4}.

\begin{figure*}[htb!]
\centering
\epsscale{1.1}
\plottwo{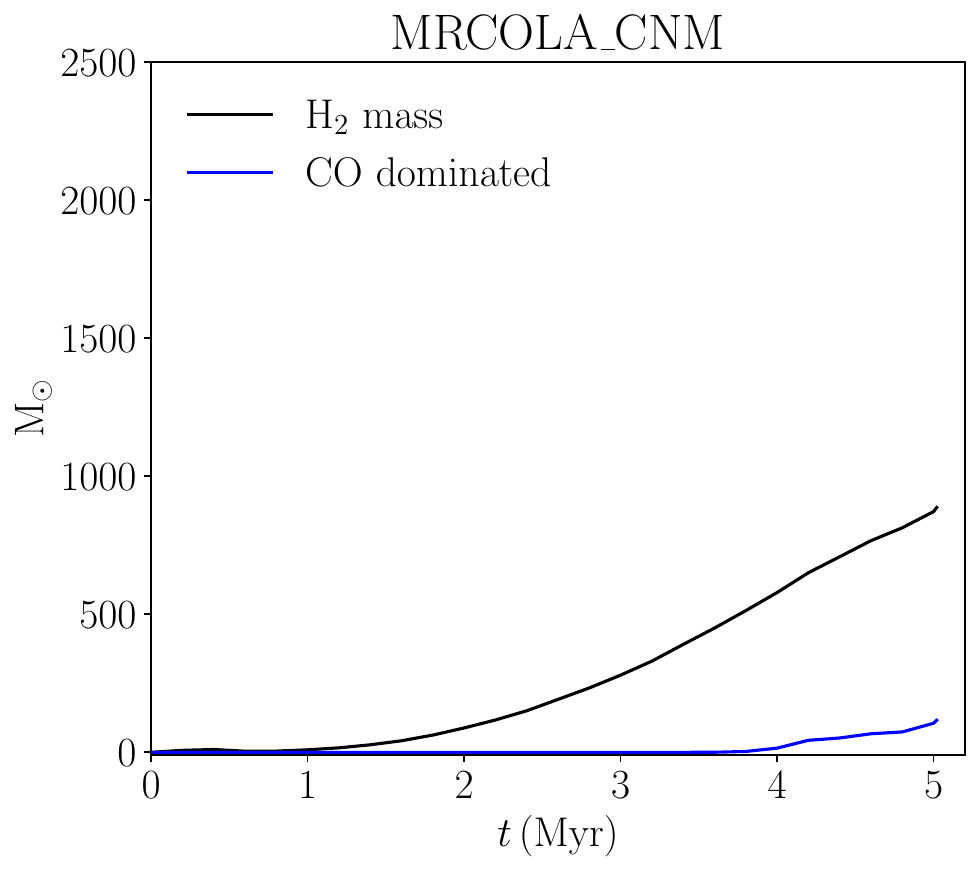}{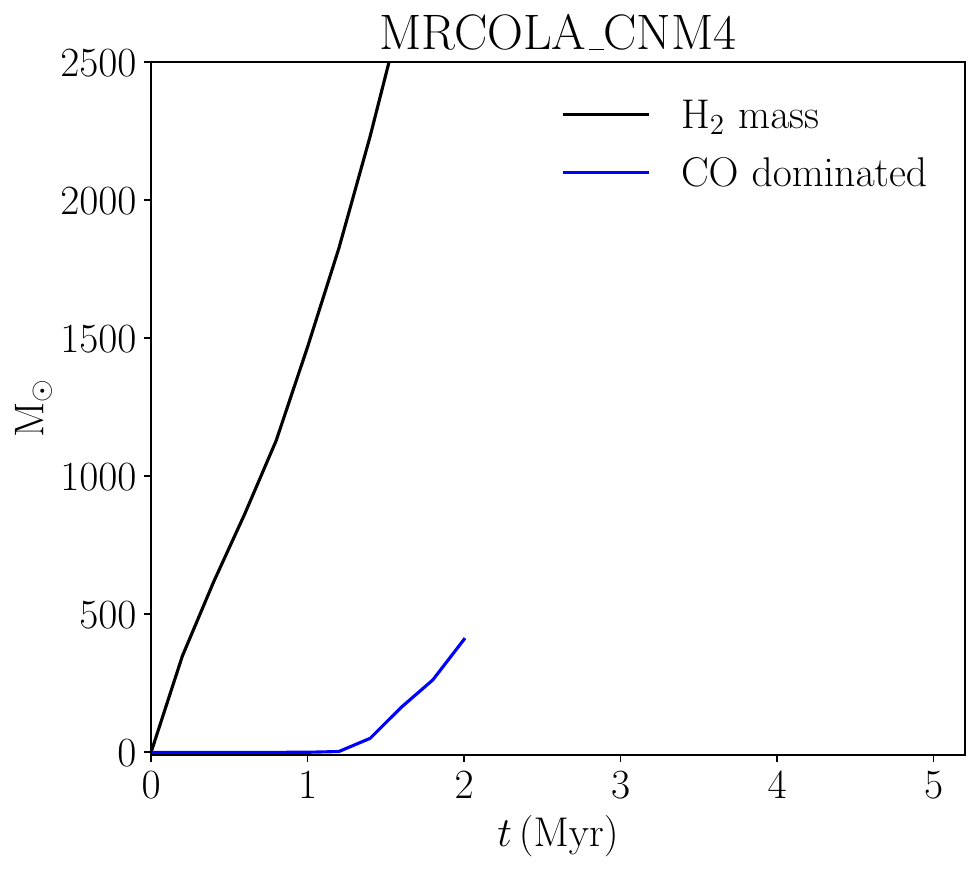}
\caption{
Time-dependent molecular gas mass comparison between
\mbox{MRCOLA\_CNM} ({\it left}) and \mbox{MRCOLA\_CNM4} ({\it right}).
The black curve shows the H$_2$ mass while the
blue curve shows the CO dominated gas mass.
The two plots have the same x-range and y-range.
\label{fig:mol}}
\end{figure*}

Figure \ref{fig:xco4} shows the [CO] for the new model. Compared to the
fiducial model (Figure \ref{fig:xco}), \mbox{MRCOLA\_CNM4} forms the
{\it filament} a factor of $\ga$2 times faster. So the starless phase
is much shorter. Moreover, the amount of molecular gas in \mbox{MRCOLA\_CNM4}
is much larger than \mbox{MRCOLA\_CNM}. Figure \ref{fig:mol} shows the
comparison. One can see the CO dominated gas in \mbox{MRCOLA\_CNM4} is
four times more than the fiducial model, which is not unexpected because
the simulation at large scales is scalable. In fact, our choices of
the initial density (30 cm$^{-3}$) and magnetic field strength (5 $\mu$G)
for the fiducial model
are challenging for dense gas formation. Our fiducial model proves that
CMR is capable of forming molecular clouds with a low initial density.
More exploration is necessary to further manifest the capability.
For example, higher collision velocity produces denser gas in
filament formation \citep[e.g.,][]{2021ApJ...916...83A,2023ApJS..265...58K}.
Note, as mentioned in \S\ref{subsec:nocmr}, the density and
field strength do not allow dense gas formation if the field is
not antiparallel. The presence of reverse fields overcome the difficulty,
highlighting an important capability of CMR.

% The main difference in the CMR-filament is that
% it can be rather isolated with no nearby major clouds, has rich, chaotic
% sub-structures (mostly fiber-like) along the {\it filament}, and most importantly,
% reverse magnetic fields on two sides of the {\it filament} regardless of the line
% of sight. More distinguishing observational features for CMR clouds shall be
% looked at in a dedicated study in the future.

\subsection{Star formation in the CMR-filament}

Since we do not consider stellar feedback in our simulations,
we stop the fiducial model simulation at 5 Myr when there are 
five sinks with a mass greater than 8 M$_\odot$. 
These massive sinks suggest that massive star formation is possible.
The massive stars would impose strong radiative feedback on the 
CMR-filament, which would change the subsequent cloud evolution.
Massive star formation in CMR-filaments can be
an interesting subject to follow up.
Additionally, protostellar outflows should also impact the {\it filament} 
\citep[e.g.][]{2010ApJ...715.1170A,2011ApJ...726...46N,2016ApJ...832...40K,2017ApJ...847..104O}.
Therefore, the discussion here about star formation in \mbox{MRCOLA\_CNM}
is in the sense of the early-stage star formation in CMR-filament.

As we can see in Figures \ref{fig:zslice} and \ref{fig:xslice},
there is no global (longitudinal) collapse of the CMR-filament through 
the end of the simulation at 5 Myr. Instead, the velocity arrows 
show that every part of the {\it filament} has inflowing gas. However,
star formation is already happening in the middle part of the {\it filament} from t=4 Myr.
By t=5 Myr, new sinks occupy a spatial span of $\sim$5 pc region along the
{\it filament}, with the possibility of a few massive stars.
These results suggest that at least in \mbox{MRCOLA\_CNM}, dense core
and star formation does not need a global cloud collapse along the long axis of the {\it filament}.
Instead the stars form within the embedded clumpy dense structures in the {\it filament}.
The {\it filament} is never a smooth cylinder throughout the simulation, but is instead 
an ensemble of the dense pieces created by the numerous 
magnetic reconnection events. The pieces are pulled together to form a
{\it filament} due to the {\it field-loop} around the {\it pancake} (Figure \ref{fig:setup}).
The whole {\it filament} and star formation process is almost a reverse of
the canonical cloud and star formation process in which a giant cloud first
forms and subsequently fragments into dense cores which possibly form new stars.
In the CMR scenario, the dense cores are not fragments of a precursor of
smooth {\it filament} but those dense pieces from supersonic 
magnetic transportation that later become gravitationally unstable.
Another example of bottom-up filamentary cloud formation is from 
\citet{2016MNRAS.455.3640S} but the physical forces are different
(one hydro+gravity while the other magnetic).

Because star formation happens before global collapse in \mbox{MRCOLA\_CNM},
the young stars are not found in a large cluster, although locally
there are signs of multiple stars in high column density regions.
As shown in \S\ref{subsec:synobs} and Figure \ref{fig:zoomcont250}, 
the new stars scatter over along the densest parts without obvious signs 
of clustering. However, it is possible that later the {\it filament} would 
develop large-scale collapse. A (mini) starburst
would happen near the collapsing center, qualitatively similar to the 
cluster formation in K22. On the other hand, a higher initial density
may result in a much denser {\it filament} \citep{2023ApJS..265...58K}
which may develop a global collapse earlier.
In reality, gravitational collapse within a large filamentary cloud is observed. 
For instance, 
the Orion A cloud \citep[with reverse magnetic fields as in][]{1997ApJS..111..245H}
is a nice analogy to our CMR-filament in \mbox{MRCOLA\_CNM}.
\citet{2017A&A...602L...2H} found evidence of a collapse toward the
OMC-1 region in Orion A using position-velocity diagrams
\citep[also see][]{2018ApJS..236...25K,2019MNRAS.489.4771G}.
The active star formation in ONC \citep{1997AJ....113.1733H}
could be fed by such collapse. %In our model \mbox{MRCOLA\_CNM},
%if a similar collapse develops later, a massive cluster can form potentially.

Regarding Orion A, we can think of its formation history the opposite way.
If Orion A were to form via a collision event but with a uniform initial
magnetic field, our control model \mbox{MRCOLA\_CNM\_sameB} showed that it
is not possible due to the subcritical mass-to-flux ratio,
unless the initial conditions are significantly different.
The field strength was based on Zeeman measurements from
\citet{1997ApJS..111..245H}, so there is probably not much leeway for
the initial field magnitude. The initial atomic gas density could be
much larger than the typical CNM density. 
Alternatively, a faster colliding
speed could also increase the dense gas production in principle,
but still the magnetic field should somehow dissipate faster.
If the initial collision was parallel to the field, then the
magnetic field should not be a problem for dense gas formation.
But it would not form a {\it filament} or explain the reverse field 
on two sides of the Orion A.
%which is the most important observational fact. 
CMR provides a compelling
case for putting together all observational pieces of 
a giant filamentary molecular cloud, most
importantly including the reverse magnetic field. 
Because a CMR-filament
naturally has a helical field, we would see reverse fields on its two 
sides regardless of the viewing angle.

\section{Summary and Conclusion}\label{sec:conclu}

In this paper, we have explored the possibility of molecular cloud 
formation through the Collision-induced Magnetic Reconnection (CMR)
mechanism in the cold neutral medium (CNM). Utilizing the \textsc{Arepo} code,
we have simulated the collision between two spherical CNM clouds in an
environment of warm neutral medium (WNM). The head-on collision happened
at the interface of a reverse magnetic field. Similar to previous studies of
CMR with molecular gas, the collision between atomic clouds triggered the CMR
process. A giant filamentary cloud formed in the collision mid-plane, with 
rich, fiber-like sub-structures across the entire 20 pc length. 
In contrast the hydrodynamic reference model formed a flattened {\it pancake} like cloud,
and the MHD case with alligned fields did not form a molecular cloud at all.
By the end of the fiducial model simulation at 5 Myr, the CMR {\it filament}
established a fully molecular (H$_2$) spine, with CO abundances
ranging from $\ga$10$^{-8}$ to $\sim$10$^{-4}$. 
Starting from 4 Myr, sinks emerged in the middle part of
the giant {\it filament}. By 5 Myr, a total of 57 sinks formed over a range of
$\sim$5 pc along the {\it filament}, 5 of which having a mass greater than
8 M$_\odot$ (most massive one reaching 16 M$_\odot$).
These results suggest that CMR with CNM is capable of forming a giant 
filamentary molecular cloud with (massive) star formation.
CMR provides a compelling case for explaining the formation of
giant filamentary molecular clouds with reverse magnetic fields (e.g., Orion A). 
In this case, magnetic fields play an active, decisive role in the {\it filament}
formation, which is different from previous ideas that the field only has a
passive role. Since CMR naturally results in a helical field around the {\it filament},
we should observe reverse magnetic fields on two sides of the {\it filament}
regardless of the viewing angle.

Synthetic observations of dust emission at far infrared wavelengths revealed
numerous dense cores along the CMR-filament. A zoom-in view of the far
infrared emission showed that many of the cores host sinks. Some of the
cores have more than one sinks. The sinks are typically located at the 
highest column density part of the core, which typically has multiple 
connecting {\it fibers} stretching out. The core-fiber structure implies that
the embedded sink is accreting through streamers, consistent with the 
findings from K22. Contrary to K22 in which the scale was a factor of 10
smaller, the sink formation in this paper is scattered over the central
5 pc length of the {\it filament} without obvious signs of clustering.
The lack of clustering is probably due to
the fact that the {\it filament} shows no sign of a global longitudinal collapse
(yet). 

The underlying logic is that CMR cloud formation does not follow
the canonical top-down picture in which a giant cloud forms first and
later fragments into dense cores that make stars. The CMR-filament
is clumpy since the beginning because reconnected {\it field-loops} bring
dense gas to the main {\it filament} piece by piece.
These merging dense pieces are created by shocks due to the
supersonic magnetic transportation. They
later becomes gravitationally unstable and form dense cores and stars.
So the CMR-filament does not need a global collapse to produce many dense
cores and stars, which reflects the chaotic nature of magnetic reconnection.
However, we do speculate that large-scale collapse along
the {\it filament} to happen later with possible star cluster formation.

CO(1-0) line emission shows again rich sub-structures in channel maps,
suggestive of turbulence in the CMR-filament.
Since the collision is head-on, the span of the velocity range reflects
the gas movement along the reconnection interface which is perpendicular 
to the collision velocity. Due to the pulling of {\it field-loops},
dense gas in the {\it pancake} moves toward the central axis with
supersonic velocities (sonic Mach number 6-10).
Therefore, the filamentary molecular cloud
formed via CMR naturally has supersonic turbulence, and the
turbulence has a magnetic origin. CMR is thus capable of
converting the coherent large-scale motion to small-scale 
turbulent motion during the formation of the {\it filament}.
Sinks are typically associated with high-velocity gas in
CO(1-0) channel maps, indicating active 
accretion toward the sinks. As long as the collision persists, the CMR
process should continuously replenish the {\it filament} with turbulence
and transfer material to the actively accreting stars,
showing a dynamic picture of cloud and star formation.

\acknowledgments 

We thank an anonymous referee for providing constructive reports. SK acknowledges fruitful discussion with Duo Xu, Hua-bai Li, Stefan Reissl, Serena Kim, and Yancy Shirley. An allocation of computer time from the UA Research Computing High Performance Computing (HPC) at the University of Arizona is gratefully acknowledged. RJS gratefully acknowledges an STFC Ernest Rutherford fellowship (grant ST/N00485X/1)  and HPC from the Durham DiRAC supercomputing facility (grants ST/P002293/1, ST/R002371/1, ST/S002502/1, and ST/R000832/1). DJW acknowledges support from the Programa de Becas Posdoctorales of the Direcci\'on General de Asuntos del Personal Acad\'{e}mico of the Universidad Nacional Aut\'{o}noma de M\'{e}xico (DGAPA,UNAM,Mexico)

\software{Astropy \citep{Astropy-Collaboration13}, Numpy \citep{numpy}, Matplotlib \citep{matplotlib}, SAOImageDS9 \citep{2003ASPC..295..489J}}

\facility{UA HPC}; 

\bibliography{ref}

\begin{thebibliography}{}
\expandafter\ifx\csname natexlab\endcsname\relax\def\natexlab#1{#1}\fi
\providecommand{\url}[1]{\href{#1}{#1}}
\providecommand{\dodoi}[1]{doi:~\href{http://doi.org/#1}{\nolinkurl{#1}}}
\providecommand{\doeprint}[1]{\href{http://ascl.net/#1}{\nolinkurl{http://ascl.net/#1}}}
\providecommand{\doarXiv}[1]{\href{https://arxiv.org/abs/#1}{\nolinkurl{https://arxiv.org/abs/#1}}}

\bibitem[{{Abe} {et~al.}(2021){Abe}, {Inoue}, {Inutsuka}, \&
  {Matsumoto}}]{2021ApJ...916...83A}
{Abe}, D., {Inoue}, T., {Inutsuka}, S.-i., \& {Matsumoto}, T. 2021, \apj, 916,
  83, \dodoi{10.3847/1538-4357/ac07a1}

\bibitem[{{Arce} {et~al.}(2010){Arce}, {Borkin}, {Goodman}, {Pineda}, \&
  {Halle}}]{2010ApJ...715.1170A}
{Arce}, H.~G., {Borkin}, M.~A., {Goodman}, A.~A., {Pineda}, J.~E., \& {Halle},
  M.~W. 2010, \apj, 715, 1170, \dodoi{10.1088/0004-637X/715/2/1170}

\bibitem[{{Astropy Collaboration} {et~al.}(2013){Astropy Collaboration},
  {Robitaille}, {Tollerud}, {Greenfield}, {Droettboom}, {Bray}, {Aldcroft},
  {Davis}, {Ginsburg}, {Price-Whelan}, {Kerzendorf}, {Conley}, {Crighton},
  {Barbary}, {Muna}, {Ferguson}, {Grollier}, {Parikh}, {Nair}, {Unther},
  {Deil}, {Woillez}, {Conseil}, {Kramer}, {Turner}, {Singer}, {Fox}, {Weaver},
  {Zabalza}, {Edwards}, {Azalee Bostroem}, {Burke}, {Casey}, {Crawford},
  {Dencheva}, {Ely}, {Jenness}, {Labrie}, {Lim}, {Pierfederici}, {Pontzen},
  {Ptak}, {Refsdal}, {Servillat}, \& {Streicher}}]{Astropy-Collaboration13}
{Astropy Collaboration}, {Robitaille}, T.~P., {Tollerud}, E.~J., {et~al.} 2013,
  \aap, 558, A33, \dodoi{10.1051/0004-6361/201322068}

\bibitem[{{Bally} {et~al.}(1987){Bally}, {Langer}, {Stark}, \&
  {Wilson}}]{1987ApJ...312L..45B}
{Bally}, J., {Langer}, W.~D., {Stark}, A.~A., \& {Wilson}, R.~W. 1987, \apjl,
  312, L45, \dodoi{10.1086/184817}

\bibitem[{{Bate} {et~al.}(1995){Bate}, {Bonnell}, \& {Price}}]{Bate95}
{Bate}, M.~R., {Bonnell}, I.~A., \& {Price}, N.~M. 1995, \mnras, 277, 362

\bibitem[{{Bhattacharjee} {et~al.}(2009){Bhattacharjee}, {Huang}, {Yang}, \&
  {Rogers}}]{2009PhPl...16k2102B}
{Bhattacharjee}, A., {Huang}, Y.-M., {Yang}, H., \& {Rogers}, B. 2009, Physics
  of Plasmas, 16, 112102, \dodoi{10.1063/1.3264103}

\bibitem[{{Burkert} \& {Hartmann}(2004)}]{2004ApJ...616..288B}
{Burkert}, A., \& {Hartmann}, L. 2004, \apj, 616, 288, \dodoi{10.1086/424895}

\bibitem[{{Chevance} {et~al.}(2023){Chevance}, {Krumholz}, {McLeod},
  {Ostriker}, {Rosolowsky}, \& {Sternberg}}]{2023ASPC..534....1C}
{Chevance}, M., {Krumholz}, M.~R., {McLeod}, A.~F., {et~al.} 2023, in
  Astronomical Society of the Pacific Conference Series, Vol. 534, Protostars
  and Planets VII, ed. S.~{Inutsuka}, Y.~{Aikawa}, T.~{Muto}, K.~{Tomida}, \&
  M.~{Tamura}, 1, \dodoi{10.48550/arXiv.2203.09570}

\bibitem[{{Clark} {et~al.}(2012{\natexlab{a}}){Clark}, {Glover}, \&
  {Klessen}}]{Clark12b}
{Clark}, P.~C., {Glover}, S.~C.~O., \& {Klessen}, R.~S. 2012{\natexlab{a}},
  \mnras, 420, 745

\bibitem[{{Clark} {et~al.}(2012{\natexlab{b}}){Clark}, {Glover}, {Klessen}, \&
  {Bonnell}}]{2012MNRAS.424.2599C}
{Clark}, P.~C., {Glover}, S. C.~O., {Klessen}, R.~S., \& {Bonnell}, I.~A.
  2012{\natexlab{b}}, \mnras, 424, 2599,
  \dodoi{10.1111/j.1365-2966.2012.21259.x}

\bibitem[{{Clark} {et~al.}(2019){Clark}, {Glover}, {Ragan}, \&
  {Duarte-Cabral}}]{Clark19}
{Clark}, P.~C., {Glover}, S. C.~O., {Ragan}, S.~E., \& {Duarte-Cabral}, A.
  2019, \mnras, 486, 4622

\bibitem[{{Draine} \& {Weingartner}(1997)}]{1997ApJ...480..633D}
{Draine}, B.~T., \& {Weingartner}, J.~C. 1997, \apj, 480, 633,
  \dodoi{10.1086/304008}

\bibitem[{{Dullemond}(2012)}]{2012ascl.soft02015D}
{Dullemond}, C.~P. 2012, {RADMC-3D: A multi-purpose radiative transfer tool},
  Astrophysics Source Code Library.
\newblock \doeprint{1202.015}

\bibitem[{{Fang} {et~al.}(2021){Fang}, {Kim}, {Pascucci}, \&
  {Apai}}]{2021ApJ...908...49F}
{Fang}, M., {Kim}, J.~S., {Pascucci}, I., \& {Apai}, D. 2021, \apj, 908, 49,
  \dodoi{10.3847/1538-4357/abcec8}

\bibitem[{{Fiege} \& {Pudritz}(2000)}]{2000MNRAS.311...85F}
{Fiege}, J.~D., \& {Pudritz}, R.~E. 2000, \mnras, 311, 85,
  \dodoi{10.1046/j.1365-8711.2000.03066.x}

\bibitem[{{Glover} {et~al.}(2010){Glover}, {Federrath}, {Mac Low}, \&
  {Klessen}}]{2010MNRAS.404....2G}
{Glover}, S.~C.~O., {Federrath}, C., {Mac Low}, M.~M., \& {Klessen}, R.~S.
  2010, \mnras, 404, 2, \dodoi{10.1111/j.1365-2966.2009.15718.x}

\bibitem[{{Goldsmith} \& {Li}(2005)}]{2005ApJ...622..938G}
{Goldsmith}, P.~F., \& {Li}, D. 2005, \apj, 622, 938, \dodoi{10.1086/428032}

\bibitem[{{Gong} {et~al.}(2017){Gong}, {Ostriker}, \& {Wolfire}}]{Gong17}
{Gong}, M., {Ostriker}, E.~C., \& {Wolfire}, M.~G. 2017, \apj, 843, 38

\bibitem[{{Gonz{\'a}lez Lobos} \& {Stutz}(2019)}]{2019MNRAS.489.4771G}
{Gonz{\'a}lez Lobos}, V., \& {Stutz}, A.~M. 2019, \mnras, 489, 4771,
  \dodoi{10.1093/mnras/stz2512}

\bibitem[{{Greif} {et~al.}(2011){Greif}, {Springel}, {White}, {Glover},
  {Clark}, {Smith}, {Klessen}, \& {Bromm}}]{Greif11}
{Greif}, T., {Springel}, V., {White}, S., {et~al.} 2011, \apj, 737, 75

\bibitem[{{Gro{\ss}schedl} {et~al.}(2018){Gro{\ss}schedl}, {Alves}, {Meingast},
  {Ackerl}, {Ascenso}, {Bouy}, {Burkert}, {Forbrich}, {F{\"u}rnkranz},
  {Goodman}, {Hacar}, {Herbst-Kiss}, {Lada}, {Larreina}, {Leschinski},
  {Lombardi}, {Moitinho}, {Mortimer}, \& {Zari}}]{2018A&A...619A.106G}
{Gro{\ss}schedl}, J.~E., {Alves}, J., {Meingast}, S., {et~al.} 2018, \aap, 619,
  A106, \dodoi{10.1051/0004-6361/201833901}

\bibitem[{{Hacar} {et~al.}(2017){Hacar}, {Alves}, {Tafalla}, \&
  {Goicoechea}}]{2017A&A...602L...2H}
{Hacar}, A., {Alves}, J., {Tafalla}, M., \& {Goicoechea}, J.~R. 2017, \aap,
  602, L2, \dodoi{10.1051/0004-6361/201730732}

\bibitem[{{Hacar} {et~al.}(2023){Hacar}, {Clark}, {Heitsch}, {Kainulainen},
  {Panopoulou}, {Seifried}, \& {Smith}}]{2023ASPC..534..153H}
{Hacar}, A., {Clark}, S.~E., {Heitsch}, F., {et~al.} 2023, in Astronomical
  Society of the Pacific Conference Series, Vol. 534, Protostars and Planets
  VII, ed. S.~{Inutsuka}, Y.~{Aikawa}, T.~{Muto}, K.~{Tomida}, \& M.~{Tamura},
  153, \dodoi{10.48550/arXiv.2203.09562}

\bibitem[{{Hacar} {et~al.}(2018){Hacar}, {Tafalla}, {Forbrich}, {Alves},
  {Meingast}, {Grossschedl}, \& {Teixeira}}]{2018A&A...610A..77H}
{Hacar}, A., {Tafalla}, M., {Forbrich}, J., {et~al.} 2018, \aap, 610, A77,
  \dodoi{10.1051/0004-6361/201731894}

\bibitem[{{Han} {et~al.}(2018){Han}, {Manchester}, {van Straten}, \&
  {Demorest}}]{2018ApJS..234...11H}
{Han}, J.~L., {Manchester}, R.~N., {van Straten}, W., \& {Demorest}, P. 2018,
  \apjs, 234, 11, \dodoi{10.3847/1538-4365/aa9c45}

\bibitem[{{Heiles}(1997)}]{1997ApJS..111..245H}
{Heiles}, C. 1997, \apjs, 111, 245, \dodoi{10.1086/313010}

\bibitem[{{Hillenbrand}(1997)}]{1997AJ....113.1733H}
{Hillenbrand}, L.~A. 1997, \aj, 113, 1733, \dodoi{10.1086/118389}

\bibitem[{{Hunter} {et~al.}(2023){Hunter}, {Clark}, {Glover}, \&
  {Klessen}}]{2023MNRAS.519.4152H}
{Hunter}, G.~H., {Clark}, P.~C., {Glover}, S. C.~O., \& {Klessen}, R.~S. 2023,
  \mnras, 519, 4152, \dodoi{10.1093/mnras/stac3751}

\bibitem[{{Hunter}(2007)}]{matplotlib}
{Hunter}, J.~D. 2007, Computing in Science and Engineering, 9, 90,
  \dodoi{10.1109/MCSE.2007.55}

\bibitem[{{Joye} \& {Mandel}(2003)}]{2003ASPC..295..489J}
{Joye}, W.~A., \& {Mandel}, E. 2003, in Astronomical Society of the Pacific
  Conference Series, Vol. 295, Astronomical Data Analysis Software and Systems
  XII, ed. H.~E. {Payne}, R.~I. {Jedrzejewski}, \& R.~N. {Hook}, 489

\bibitem[{{Kong}(2022)}]{2022ApJ...933...40K}
{Kong}, S. 2022, \apj, 933, 40, \dodoi{10.3847/1538-4357/ac70cd}

\bibitem[{{Kong} {et~al.}(2019){Kong}, {Arce}, {Maureira}, {Caselli}, {Tan}, \&
  {Fontani}}]{2019ApJ...874..104K}
{Kong}, S., {Arce}, H.~G., {Maureira}, M.~J., {et~al.} 2019, \apj, 874, 104,
  \dodoi{10.3847/1538-4357/ab07b9}

\bibitem[{{Kong} {et~al.}(2023){Kong}, {Ossenkopf-Okada}, {Arce}, {Klessen}, \&
  {Xu}}]{2023ApJS..265...58K}
{Kong}, S., {Ossenkopf-Okada}, V., {Arce}, H.~G., {Klessen}, R.~S., \& {Xu}, D.
  2023, \apjs, 265, 58, \dodoi{10.3847/1538-4365/acbfb0}

\bibitem[{{Kong} {et~al.}(2022){Kong}, {Whitworth}, {Smith}, \&
  {Hamden}}]{2022MNRAS.517.4679K}
{Kong}, S., {Whitworth}, D.~J., {Smith}, R.~J., \& {Hamden}, E.~T. 2022,
  \mnras, 517, 4679, \dodoi{10.1093/mnras/stac2932}

\bibitem[{{Kong} {et~al.}(2018){Kong}, {Arce}, {Feddersen}, {Carpenter},
  {Nakamura}, {Shimajiri}, {Isella}, {Ossenkopf-Okada}, {Sargent},
  {S{\'a}nchez-Monge}, {Suri}, {Kauffmann}, {Pillai}, {Pineda}, {Koda},
  {Bally}, {Lis}, {Padoan}, {Klessen}, {Mairs}, {Goodman}, {Goldsmith},
  {McGehee}, {Schilke}, {Teuben}, {Maureira}, {Hara}, {Ginsburg}, {Burkhart},
  {Smith}, {Schmiedeke}, {Pineda}, {Ishii}, {Sasaki}, {Kawabe}, {Urasawa},
  {Oyamada}, \& {Tanabe}}]{2018ApJS..236...25K}
{Kong}, S., {Arce}, H.~G., {Feddersen}, J.~R., {et~al.} 2018, \apjs, 236, 25,
  \dodoi{10.3847/1538-4365/aabafc}

\bibitem[{{Kong} {et~al.}(2021){Kong}, {Ossenkopf-Okada}, {Arce}, {Bally},
  {S{\'a}nchez-Monge}, {McGehee}, {Suri}, {Klessen}, {Carpenter}, {Lis},
  {Nakamura}, {Schilke}, {Smith}, {Mairs}, {Goodman}, \&
  {Maureira}}]{2021ApJ...906...80K}
{Kong}, S., {Ossenkopf-Okada}, V., {Arce}, H.~G., {et~al.} 2021, \apj, 906, 80,
  \dodoi{10.3847/1538-4357/abc687}

\bibitem[{{Kuiper} {et~al.}(2016){Kuiper}, {Turner}, \&
  {Yorke}}]{2016ApJ...832...40K}
{Kuiper}, R., {Turner}, N.~J., \& {Yorke}, H.~W. 2016, \apj, 832, 40,
  \dodoi{10.3847/0004-637X/832/1/40}

\bibitem[{{Lada} {et~al.}(2009){Lada}, {Lombardi}, \&
  {Alves}}]{2009ApJ...703...52L}
{Lada}, C.~J., {Lombardi}, M., \& {Alves}, J.~F. 2009, \apj, 703, 52,
  \dodoi{10.1088/0004-637X/703/1/52}

\bibitem[{{Lazarian} \& {Hoang}(2007)}]{2007MNRAS.378..910L}
{Lazarian}, A., \& {Hoang}, T. 2007, \mnras, 378, 910,
  \dodoi{10.1111/j.1365-2966.2007.11817.x}

\bibitem[{{Lewis} {et~al.}(2021){Lewis}, {Lada}, {Bieging}, {Kazarians},
  {Alves}, \& {Lombardi}}]{2021ApJ...908...76L}
{Lewis}, J.~A., {Lada}, C.~J., {Bieging}, J., {et~al.} 2021, \apj, 908, 76,
  \dodoi{10.3847/1538-4357/abc41f}

\bibitem[{{Li} \& {Goldsmith}(2003)}]{2003ApJ...585..823L}
{Li}, D., \& {Goldsmith}, P.~F. 2003, \apj, 585, 823, \dodoi{10.1086/346227}

\bibitem[{{Li} \& {Goldsmith}(2012)}]{2012ApJ...756...12L}
---. 2012, \apj, 756, 12, \dodoi{10.1088/0004-637X/756/1/12}

\bibitem[{{Li} \& {Nakamura}(2004)}]{2004ApJ...609L..83L}
{Li}, Z.-Y., \& {Nakamura}, F. 2004, \apjl, 609, L83, \dodoi{10.1086/422839}

\bibitem[{{Mathis} {et~al.}(1977){Mathis}, {Rumpl}, \&
  {Nordsieck}}]{1977ApJ...217..425M}
{Mathis}, J.~S., {Rumpl}, W., \& {Nordsieck}, K.~H. 1977, \apj, 217, 425,
  \dodoi{10.1086/155591}

\bibitem[{{Nakamura} {et~al.}(2011){Nakamura}, {Kamada}, {Kamazaki}, {Kawabe},
  {Kitamura}, {Shimajiri}, {Tsukagoshi}, {Tachihara}, {Akashi}, {Azegami},
  {Ikeda}, {Kurono}, {Li}, {Miura}, {Nishi}, \&
  {Umemoto}}]{2011ApJ...726...46N}
{Nakamura}, F., {Kamada}, Y., {Kamazaki}, T., {et~al.} 2011, \apj, 726, 46,
  \dodoi{10.1088/0004-637X/726/1/46}

\bibitem[{{Nakano} \& {Nakamura}(1978)}]{1978PASJ...30..671N}
{Nakano}, T., \& {Nakamura}, T. 1978, \pasj, 30, 671

\bibitem[{{Nelson} \& {Langer}(1999)}]{1999ApJ...524..923N}
{Nelson}, R.~P., \& {Langer}, W.~D. 1999, \apj, 524, 923,
  \dodoi{10.1086/307823}

\bibitem[{{Offner} \& {Chaban}(2017)}]{2017ApJ...847..104O}
{Offner}, S. S.~R., \& {Chaban}, J. 2017, \apj, 847, 104,
  \dodoi{10.3847/1538-4357/aa8996}

\bibitem[{{Ossenkopf} \& {Henning}(1994)}]{1994A&A...291..943O}
{Ossenkopf}, V., \& {Henning}, T. 1994, \aap, 291, 943

\bibitem[{{Pakmor} {et~al.}(2011){Pakmor}, {Bauer}, \& {Springel}}]{Pakmor11}
{Pakmor}, R., {Bauer}, A., \& {Springel}, V. 2011, \mnras, 418, 1392

\bibitem[{{Palmeirim} {et~al.}(2013){Palmeirim}, {Andr{\'e}}, {Kirk},
  {Ward-Thompson}, {Arzoumanian}, {K{\"o}nyves}, {Didelon}, {Schneider},
  {Benedettini}, {Bontemps}, {Di Francesco}, {Elia}, {Griffin}, {Hennemann},
  {Hill}, {Martin}, {Men'shchikov}, {Molinari}, {Motte}, {Nguyen Luong},
  {Nutter}, {Peretto}, {Pezzuto}, {Roy}, {Rygl}, {Spinoglio}, \&
  {White}}]{2013A&A...550A..38P}
{Palmeirim}, P., {Andr{\'e}}, P., {Kirk}, J., {et~al.} 2013, \aap, 550, A38,
  \dodoi{10.1051/0004-6361/201220500}

\bibitem[{{Poidevin} {et~al.}(2011){Poidevin}, {Bastien}, \&
  {Jones}}]{2011ApJ...741..112P}
{Poidevin}, F., {Bastien}, P., \& {Jones}, T.~J. 2011, \apj, 741, 112,
  \dodoi{10.1088/0004-637X/741/2/112}

\bibitem[{{Polychroni} {et~al.}(2013){Polychroni}, {Schisano}, {Elia}, {Roy},
  {Molinari}, {Martin}, {Andr{\'e}}, {Turrini}, {Rygl}, {Di Francesco},
  {Benedettini}, {Busquet}, {di Giorgio}, {Pestalozzi}, {Pezzuto},
  {Arzoumanian}, {Bontemps}, {Hennemann}, {Hill}, {K{\"o}nyves},
  {Men'shchikov}, {Motte}, {Nguyen-Luong}, {Peretto}, {Schneider}, \&
  {White}}]{2013ApJ...777L..33P}
{Polychroni}, D., {Schisano}, E., {Elia}, D., {et~al.} 2013, \apjl, 777, L33,
  \dodoi{10.1088/2041-8205/777/2/L33}

\bibitem[{{Reissl} {et~al.}(2021){Reissl}, {Stutz}, {Klessen}, {Seifried}, \&
  {Walch}}]{2021MNRAS.500..153R}
{Reissl}, S., {Stutz}, A.~M., {Klessen}, R.~S., {Seifried}, D., \& {Walch}, S.
  2021, \mnras, 500, 153, \dodoi{10.1093/mnras/staa3148}

\bibitem[{{Reissl} {et~al.}(2016){Reissl}, {Wolf}, \&
  {Brauer}}]{2016A&A...593A..87R}
{Reissl}, S., {Wolf}, S., \& {Brauer}, R. 2016, \aap, 593, A87,
  \dodoi{10.1051/0004-6361/201424930}

\bibitem[{{Sembach} {et~al.}(2000){Sembach}, {Howk}, {Ryans}, \&
  {Keenan}}]{2000ApJ...528..310S}
{Sembach}, K.~R., {Howk}, J.~C., {Ryans}, R. S.~I., \& {Keenan}, F.~P. 2000,
  \apj, 528, 310, \dodoi{10.1086/308173}

\bibitem[{{Shimajiri} {et~al.}(2019){Shimajiri}, {Andr{\'e}}, {Palmeirim},
  {Arzoumanian}, {Bracco}, {K{\"o}nyves}, {Ntormousi}, \&
  {Ladjelate}}]{2019A&A...623A..16S}
{Shimajiri}, Y., {Andr{\'e}}, P., {Palmeirim}, P., {et~al.} 2019, \aap, 623,
  A16, \dodoi{10.1051/0004-6361/201834399}

\bibitem[{{Smith} {et~al.}(2016){Smith}, {Glover}, {Klessen}, \&
  {Fuller}}]{2016MNRAS.455.3640S}
{Smith}, R.~J., {Glover}, S.~C.~O., {Klessen}, R.~S., \& {Fuller}, G.~A. 2016,
  \mnras, 455, 3640, \dodoi{10.1093/mnras/stv2559}

\bibitem[{{Smith} {et~al.}(2020){Smith}, {Tre{\ss}}, {Sormani}, {Glover},
  {Klessen}, {Clark}, {Izquierdo}, {Duarte-Cabral}, \&
  {Zucker}}]{2020MNRAS.492.1594S}
{Smith}, R.~J., {Tre{\ss}}, R.~G., {Sormani}, M.~C., {et~al.} 2020, \mnras,
  492, 1594, \dodoi{10.1093/mnras/stz3328}

\bibitem[{{Soler}(2019)}]{2019A&A...629A..96S}
{Soler}, J.~D. 2019, \aap, 629, A96, \dodoi{10.1051/0004-6361/201935779}

\bibitem[{{Soler} {et~al.}(2018){Soler}, {Bracco}, \&
  {Pon}}]{2018A&A...609L...3S}
{Soler}, J.~D., {Bracco}, A., \& {Pon}, A. 2018, \aap, 609, L3,
  \dodoi{10.1051/0004-6361/201732203}

\bibitem[{{Springel}(2005)}]{Springel05}
{Springel}, V. 2005, \mnras, 364, 1105

\bibitem[{{Springel}(2010)}]{Springel10}
---. 2010, \mnras, 401, 791

\bibitem[{{Stephens} {et~al.}(2017){Stephens}, {Dunham}, {Myers}, {Pokhrel},
  {Sadavoy}, {Vorobyov}, {Tobin}, {Pineda}, {Offner}, {Lee}, {Kristensen},
  {J{\o}rgensen}, {Goodman}, {Bourke}, {Arce}, \&
  {Plunkett}}]{2017ApJ...846...16S}
{Stephens}, I.~W., {Dunham}, M.~M., {Myers}, P.~C., {et~al.} 2017, \apj, 846,
  16, \dodoi{10.3847/1538-4357/aa8262}

\bibitem[{{Strittmatter}(1966)}]{1966MNRAS.132..359S}
{Strittmatter}, P.~A. 1966, \mnras, 132, 359, \dodoi{10.1093/mnras/132.2.359}

\bibitem[{{Stutz} \& {Kainulainen}(2015)}]{2015A&A...577L...6S}
{Stutz}, A.~M., \& {Kainulainen}, J. 2015, \aap, 577, L6,
  \dodoi{10.1051/0004-6361/201526243}

\bibitem[{{Tahani} {et~al.}(2018){Tahani}, {Plume}, {Brown}, \&
  {Kainulainen}}]{2018A&A...614A.100T}
{Tahani}, M., {Plume}, R., {Brown}, J.~C., \& {Kainulainen}, J. 2018, \aap,
  614, A100, \dodoi{10.1051/0004-6361/201732219}

\bibitem[{{Tahani} {et~al.}(2019){Tahani}, {Plume}, {Brown}, {Soler}, \&
  {Kainulainen}}]{2019A&A...632A..68T}
{Tahani}, M., {Plume}, R., {Brown}, J.~C., {Soler}, J.~D., \& {Kainulainen}, J.
  2019, \aap, 632, A68, \dodoi{10.1051/0004-6361/201936280}

\bibitem[{{Tahani} {et~al.}(2022){Tahani}, {Lupypciw}, {Glover}, {Plume},
  {West}, {Kothes}, {Inutsuka}, {Lee}, {Robishaw}, {Knee}, {Brown}, {Doi},
  {Grenier}, \& {Haverkorn}}]{2022A&A...660A..97T}
{Tahani}, M., {Lupypciw}, W., {Glover}, J., {et~al.} 2022, \aap, 660, A97,
  \dodoi{10.1051/0004-6361/202141170}

\bibitem[{{Tang} {et~al.}(2020){Tang}, {Zuo}, {Li}, {Qian}, {Liu}, {Wu},
  {Kr{\v{c}}o}, {Liu}, {Yue}, {Zhu}, {Liu}, {Yu}, {Sun}, {Jiang}, {Pan}, {Li},
  {Gan}, {Yao}, {Liu}, \& {FAST Collaboration}}]{2020RAA....20...77T}
{Tang}, N.-Y., {Zuo}, P., {Li}, D., {et~al.} 2020, Research in Astronomy and
  Astrophysics, 20, 077, \dodoi{10.1088/1674-4527/20/5/77}

\bibitem[{{Tielens} \& {Hollenbach}(1985)}]{1985ApJ...291..722T}
{Tielens}, A.~G.~G.~M., \& {Hollenbach}, D. 1985, \apj, 291, 722,
  \dodoi{10.1086/163111}

\bibitem[{{Tress} {et~al.}(2020){Tress}, {Smith}, {Sormani}, {Glover},
  {Klessen}, {Mac Low}, \& {Clark}}]{2020MNRAS.492.2973T}
{Tress}, R.~G., {Smith}, R.~J., {Sormani}, M.~C., {et~al.} 2020, \mnras, 492,
  2973, \dodoi{10.1093/mnras/stz3600}

\bibitem[{{Truelove} {et~al.}(1997){Truelove}, {Klein}, {McKee}, {Holliman},
  {Howell}, \& {Greenough}}]{Truelove97}
{Truelove}, J.~K., {Klein}, R.~I., {McKee}, C.~F., {et~al.} 1997, \apjl, 489,
  L179

\bibitem[{van~der Walt {et~al.}(2011)van~der Walt, Colbert, \&
  Varoquaux}]{numpy}
van~der Walt, S., Colbert, S.~C., \& Varoquaux, G. 2011, Computing in Science
  \& Engineering, 13, 22, \dodoi{10.1109/MCSE.2011.37}

\bibitem[{{V{\'a}zquez-Semadeni} {et~al.}(2019){V{\'a}zquez-Semadeni}, {Palau},
  {Ballesteros-Paredes}, {G{\'o}mez}, \&
  {Zamora-Avil{\'e}s}}]{2019MNRAS.490.3061V}
{V{\'a}zquez-Semadeni}, E., {Palau}, A., {Ballesteros-Paredes}, J.,
  {G{\'o}mez}, G.~C., \& {Zamora-Avil{\'e}s}, M. 2019, \mnras, 490, 3061,
  \dodoi{10.1093/mnras/stz2736}

\bibitem[{{Wolfire} {et~al.}(2010){Wolfire}, {Hollenbach}, \&
  {McKee}}]{2010ApJ...716.1191W}
{Wolfire}, M.~G., {Hollenbach}, D., \& {McKee}, C.~F. 2010, \apj, 716, 1191,
  \dodoi{10.1088/0004-637X/716/2/1191}

\bibitem[{{Wu} {et~al.}(2020){Wu}, {Tan}, {Christie}, \&
  {Nakamura}}]{2020ApJ...891..168W}
{Wu}, B., {Tan}, J.~C., {Christie}, D., \& {Nakamura}, F. 2020, \apj, 891, 168,
  \dodoi{10.3847/1538-4357/ab77b5}

\bibitem[{{Zhao} {et~al.}(2024){Zhao}, {Pudritz}, {Pillsworth}, {Robinson}, \&
  {Wadsley}}]{2024arXiv240518474Z}
{Zhao}, B., {Pudritz}, R.~E., {Pillsworth}, R., {Robinson}, H., \& {Wadsley},
  J. 2024, arXiv e-prints, arXiv:2405.18474, \dodoi{10.48550/arXiv.2405.18474}

\end{thebibliography}
\bibliographystyle{aasjournal}

\end{document}